\documentclass[prd,aps,showpacs,notitlepage,superscriptaddress,nofootinbib]{revtex4-1}

\usepackage{amsmath}
\usepackage{multirow}
\usepackage{float}
\usepackage{soul}
\usepackage{caption}
\usepackage {tikz}

\def\bea#1\eea{\begin{align}#1\end{align}}

\makeatletter
\def\slash#1{{\mathpalette\c@ncel{#1}}} 
\makeatother

\newcommand\beq{\begin{eqnarray}}
\newcommand\eeq{\end{eqnarray}}

\newcommand\la{\langle}
\newcommand\ra{\rangle}

\begin{document}
\title{Seeking a possible approach to understanding the origin of the proton mass using 
twist-4 contributions in inclusive reactions}

\date{\today}

\author{Xiaohui Wen}
\email{xiaohuiwen1202@outlook.com}
\affiliation{State Key Laboratory of Nuclear Physics and Technology, Institute of
Quantum Matter, South China Normal University, Guangzhou 510006, China}
\affiliation{Key Laboratory of Atomic and Subatomic Structure and
Quantum Control (MOE), Guangdong-Hong Kong Joint Laboratory
of Quantum Matter, Guangzhou 510006, China}
\affiliation{Guangdong Basic Research Center of Excellence for
Structure and Fundamental Interactions of Matter, Guangdong
Provincial Key Laboratory of Nuclear Science, Guangzhou
510006, China}

\author{Shinsuke Yoshida}
\email{shinyoshida85@gmail.com}
\affiliation{State Key Laboratory of Nuclear Physics and Technology, Institute of
Quantum Matter, South China Normal University, Guangzhou 510006, China}
\affiliation{Guangdong Basic Research Center of Excellence for
Structure and Fundamental Interactions of Matter, Guangdong
Provincial Key Laboratory of Nuclear Science, Guangzhou
510006, China}

\begin{abstract}

Most of the mass of the visible universe originates from the strong interaction.
Quantitatively reproducing the origin of the nucleon mass is one of the important challenges in 
fundamental science. 
Since the next-generation Electron-Ion Collider has identified resolving this problem as one of its 
major goals, further theoretical studies are needed to develop approaches to investigating 
the origin experimentally. 
In this paper, we explore the possibility of evaluating the
``gluon condensate term'' which is particularly difficult to determine among the contributions to the 
nucleon mass, using high-energy scattering experiments. We first express the gluon condensate
 term in terms of experimentally accessible twist-4 lightcone matrix elements. We then derive,
  for the first time, analytic formulas for the cross section including all twist-4 contributions 
  from pure quark, quark-gluon, and pure gluon matrix elements in $ep$ collisions, 
  and show how these contributions can be related to the gluon condensate term.

\end{abstract}

\maketitle


\section{Introduction}

Understanding the origin of the proton mass is one of the fundamental challenges in quantum chromodynamics (QCD). Although the current quark masses account for only a small fraction of 
the proton mass, the remaining mass is generated dynamically through the strong interaction. 
The QCD energy-momentum tensor provides a theoretical framework for investigating this problem. 
In particular, its trace is given by
\beq
T^\mu_{\ \mu}=\frac{\beta(g)}{2g}F_a^{\mu\nu}F_{a\mu\nu}
+(1+\gamma_m)\sum_q m_q\bar{\psi}_q\psi_q,
\eeq
which receives contributions from the QCD trace anomaly\cite{Collins:1976yq}. 
The proton matrix element of this gluonic operator, which we refer to as the gluon condensate term, is 
an important quantity for understanding the origin of the proton mass\cite{Ji:1994av}. 
Although lattice QCD has made significant progress in the proton mass decomposition\cite{Yang:2018nqn}, 
a precise direct evaluation of the gluon condensate term remains challenging.

Understanding the origin of the proton mass is also one of the important physics goals of the future 
Electron-Ion Collider(EIC)\cite{EIC_white}. 
Several experimental approaches have been proposed to investigate the 
gluonic structure associated with the proton mass. In particular, near-threshold $J/\psi$ photoproduction
has attracted considerable attention as a possible probe of the gluonic gravitational form factors 
and the QCD trace anomaly\cite{Hatta:2018ina,GlueX:2019mkq,GlueX:2023pev}. 
However, the extraction of the gluon condensate term from this process 
is also challenging because of theoretical uncertainties and experimental difficulties.
It is therefore important to explore other possibilities for experimentally determining this quantity.

In this paper, we investigate a possible approach based on twist-4 contributions in high-energy inclusive
reactions. Since $F_a^{\mu\nu}F_{a\mu\nu}$ is a local twist-4 operator, it is natural to consider whether
its proton matrix element can be expressed in terms of twist-4 lightcone distribution functions. Previous
studies have investigated the relation between the gluon condensate term and lightcone 
distributions\cite{Ji:2020baz,Hatta:2020iin}.
We further develop this approach by expressing the gluon condensate term in terms of multiparton
correlation functions which can contribute to experimentally measurable cross sections. If these functions
can be constrained by high-energy scattering experiments, they may provide an alternative way to
obtain information on the gluon condensate term.

The theoretical framework for calculating higher twist contributions has been extensively developed in the
context of transverse spin asymmetries. In particular, the collinear twist-3 formalism has been successfully
applied to various hard scattering processes and systematic methods based on the collinear expansion
and Ward-Takahashi identities have been established to obtain gauge-invariant cross-section 
formulas\cite{Eguchi:2006mc,Beppu:2010qn,Kanazawa:2013uia,Xing:2019ovj,Yoshida:2022vnf,Ikarashi:2022zeo}. 
Furthermore, the importance of identifying a complete set of independent functions using the equations of
motion and translational invariance has been recognized from the viewpoint of gauge and boost invariance 
of cross sections\cite{Kanazawa:2015ajw}.
Building on the developments in twist-3 studies, extensions to twist-4 have been investigated in various
research areas\cite{Qiu:1988dn,Bartels:1999km,Guo:2001tz,Xing:2011fb,Kang:2013ufa,Kang:2013raa,Kang:2014ela,Kang:2016ron,Fu:2023jqv,Fu:2024sba}. 
However, previous calculations have been limited to specific contributions, such as 
double pole contributions, and no formulation that systematically incorporates all contributions has yet 
been established.
In particular, a unified treatment of pure quark, quark-gluon, and pure gluon contributions is necessary for 
our purpose. The techniques developed in twist-3 studies provide a useful basis for extending the
diagrammatic method to twist-4 and deriving the complete contributions associated with four-parton 
matrix elements.

In this paper, we first express the gluon condensate term in terms of twist-4 lightcone matrix
elements using techniques developed in twist-3 studies. We show that these matrix elements can be 
rewritten in terms of pure-quark, quark-gluon, and pure-gluon correlation functions,
some of which are also related to transverse-momentum-dependent(TMD) 
distributions\cite{Boussarie:2023izj}.
We then apply the diagrammatic method\cite{Ellis:1982wd,Ellis:1982cd} 
to derive the twist-4 contributions from four-operator 
matrix elements to unpolarized cross sections in electron-proton collisions. By systematically
using the Ward-Takahashi identities, we obtain gauge-invariant expressions and demonstrate 
that the matrix elements appearing in the gluon condensate term also contribute to 
cross sections inclusive scattering. Our results provide a theoretical basis for exploring the possibility 
of determining the gluon condensate term through high energy scattering
experiments.

The remainder of this paper is organized as follows.
In Sec. II, we express the gluon condensate term in terms of twist-4 lightcone matrix elements
and show how these matrix elements can be rewritten in terms of experimentally accessible
twist-4 matrix elements. We also discuss their relations to TMD matrix elements.
In Sec. III, we derive the quark-gluon and 4-quark twist-4 contributions to the
unpolarized cross section for light-hadron production in $ep$ collisions
using the diagrammatic method based on the Ward-Takahashi identities.
In Sec. IV, we extend this method to the pure gluon twist-4 contributions
to the unpolarized cross section in semi-inclusive deep-inelastic scattering.
Sec. V is devoted to a summary of our study.


\section{Gluon condensate term in terms of lightcone twist-4 distribution matrix elements}

We first express the gluon condensate term $\la P|F_a^{\mu\nu}(0)F_{a\mu\nu}(0)|P\ra$
in terms of twist-4 lightcone distribution matrix elements which could be measured in
experiments.
Although there have been previous studies along these line\cite{Ji:2020baz,Hatta:2020iin}, 
we revisit the derivation to clarify its connection to cross sections in inclusive processes 
discussed in this paper. We use two lightlike vectors $p$ and $n$ which
satisfy $p\cdot n=p^+n^-=1$ for convenience.
The gluon condensate term is expanded as
\beq
\la P|F_a^{\mu\nu}(0)F_a^{\mu\nu}(0)|P\ra
=\la P|F_a^{ij}(0)F_a^{ij}(0)|P\ra
-2\la P|F_a^{pn}(0)F_a^{pn}(0)|P\ra
+4\la P|F_a^{in}(0)F_a^{pi}(0)|P\ra.
\label{gluon_condensate}
\eeq
We use the shorthand notation $F_a^{pn}(0)=F_a^{\alpha\beta}(0)p_{\alpha}n_{\beta}$
throughout this paper. We start with the calculation of the first term 
$\la P|F_a^{ij}(0)F_a^{ij}(0)|P\ra$.
We rewrite it in terms of the lightcone matrix element,
\beq
T_1&=&\la P|F_a^{ij}(0)F_a^{ij}(0)|P\ra
=\int dx\int{d\lambda\over 2\pi}\,e^{i\lambda x}
\la P|F_b^{ij}(0)[0,\lambda n]_{ba}F_a^{ij}(\lambda n)|P\ra,
\eeq
where $[0,\lambda]_{ba}$ represents the Wilson line in the adjoint representation,
\beq
[0,\lambda]_{ba}={\rm P}\Bigl(ig\int^{0}_{\lambda}dt\,A_c^n(tn)T^c\Bigr)_{ba},\qquad
(T^{c})_{ba}=-if^{cba}.
\label{Wilson_ad}
\eeq
We do not write adjoint indices explicitly unless necessary.
Integrating by parts with respect to $\lambda$, we can introduce covariant derivatives
into the matrix elements
\beq
&&-{1\over x^2}\int{d\lambda\over 2\pi}\,{\partial^2\over \partial\lambda^2}e^{i\lambda x}
\la P|F^{ij}(0)[0,\lambda n]F^{ij}(\lambda n)|P\ra
={1\over x^2}\int{d\lambda\over 2\pi}\,e^{i\lambda x}\la P|F^{ij}(0)\overleftarrow{D}^n(0)
[0,\lambda n]D^n(\lambda n)F^{ij}(\lambda n)|P\ra\Bigr].
\eeq
Note that Ref. \cite{Hatta:2020iin} suggests that the $x$-dependent distribution 
contains a $\delta(x)$ term. In the present approach, we will focus only on calculating the finite part
without this delta function term.
Using the Bianchi identity
\beq
D^{\mu}F^{\nu\lambda}+D^{\nu}F^{\lambda\mu}+D^{\lambda}F^{\mu\nu}=0,
\label{Bianchi}
\eeq
we can show
\beq
T_1&=&2\int {dx\over x^2}\Bigl\{\int{d\lambda\over 2\pi}e^{i\lambda x}\Bigl[
\la P|F^{jn}(0)\overleftarrow{D}_{\perp}^i(0)[0,\lambda n]
D_{\perp}^i(\lambda n)F^{j n}(\lambda n)|P\ra
-\la P|F^{in}(0)\overleftarrow{D}_{\perp}^{j}(0)[0,\lambda n]
D_{\perp}^i(\lambda n)F^{j n}(\lambda n)|P\ra
\Bigr]\Bigr\}
\nonumber\\
&=&-2\int {dx\over x^2}\Bigl\{
(g_{\perp\mu\nu}g_{\perp\alpha\beta}-g_{\perp\mu\beta}g_{\perp\alpha\nu})
\int{d\lambda\over 2\pi}e^{i\lambda x}\Bigl[
\la P|F^{\nu n}(0)[0,\lambda n]D_{\perp}^{\beta}(\lambda n)
D_{\perp}^{\alpha}(\lambda n)F^{\mu n}(\lambda n)|P\ra
\nonumber\\
&&+i\int^{0}_{\lambda}
d\mu\,\la P|F^{\nu n}(0)[0,\mu n]gF^{\beta n}(\mu n)[\mu n,\lambda n]
D_{\perp}^{\alpha}(\lambda n)F^{\mu n}(\lambda n)|P\ra
\Bigr]\Bigr\},
\label{T_1}
\eeq
where we used the translation invariance of the forward matrix element\cite{Kanazawa:2015ajw},
\beq
&&\la P|F^{jn}(0)\overleftarrow{D}_{\perp}^i(0)[0,\lambda n]
D_{\perp}^i(\lambda n)F^{j n}(\lambda n)|P\ra
+\la P|F^{jn}(0)[0,\lambda n]
D_{\perp}^i(\lambda n)D_{\perp}^i(\lambda n)F^{j n}(\lambda n)|P\ra
\nonumber\\
&&+i\int^0_{\lambda}d\mu\,\la P|F^{jn}(0)[0,\mu n]gF^{i n}(\mu n)[\mu n,\lambda n]
D_{\perp}^i(\lambda n)F^{j n}(\lambda n)|P\ra=0.
\eeq
The second term in (\ref{gluon_condensate}) is calculated in the same way,
\beq
T_2&=&-2\la P|F^{pn}(0)F^{pn}(0)|P\ra
=\int dx\Big[-{2\over x^2}\int{d\lambda\over 2\pi}\,e^{i\lambda x}\la P|F^{pn}(0)\overleftarrow{D}^n(0)
[0,\lambda n]D^n(\lambda n)F^{pn}(\lambda n)|P\ra\Bigr].
\eeq
We use the equations of motion
\beq
D_{ab\,\alpha}F_b^{\alpha n}=D_{ab}^nF_b^{p n}+D_{ab\,\perp\alpha}F_b^{\alpha n}
=-g\bar{\psi}t^a\slash{n}\psi,
\label{EOM}
\eeq
where $t^a$ is a generator of SU($N_c$) group.
Thus $T_2$ is rewritten as
\beq
T_2&=&-2\int {dx\over x^2}\Bigl\{\int{d\lambda\over 2\pi}e^{i\lambda x}
\Bigl[
-\la P|F^{i n}(0)[0,\lambda n]D^i_{\perp}(\lambda n)
D_{\perp}^j(\lambda n)F^{j n}(\lambda n)|P\ra
\nonumber\\
&&-ig\int^{0}_{\lambda}d\omega\,\la P|F^{i n}(0)[0,\omega n]
F^{i n}(\omega n)[\omega n,\lambda n]
D^j_{\perp}(\lambda n)F^{j n}(\lambda n)|P\ra
\nonumber\\
&&+2\la P|\bar{\psi}(0)\{0,\lambda n\}g\Bigl(D_{\perp i}(\lambda n)F^{i n}(\lambda n)\Big)
\{\lambda n,0\}\slash{n}\psi(0)|P\ra
\nonumber\\
&&-{g^2\over 2}(\slash{n})_{ji}(\slash{n})_{lk}
\la P|\bar{\psi}_j(0)\{0,\lambda n\}\psi_k(\lambda n)\bar{\psi}_l(\lambda n)\{\lambda n,0\}\psi_i(0)|P\ra
-{g^2\over 2N_c}\la P|\bar{\psi}(0)\slash{n}\psi(0)\bar{\psi}(\lambda n)\slash{n}\psi(\lambda n)|P\ra
\Bigr]\Bigr\},
\label{T_2}
\eeq
where $\Bigl(D_{\perp i}F^{i n}\Bigr)
=t^aD_{\perp ab\,i}F_b^{i n}$
and $\{...\}$ represents the Wilson line in the fundamental representation given by changing
$T^{c}$ to $t^c$ in (\ref{Wilson_ad}).
We used the formula $[0,\lambda n]_{ba}t^a=\{\lambda n,0\}t^b\{0,\lambda n\}$,
the Fierz identity
\beq
(t^a)_{ij}(t^{a})_{kl}={1\over 2}\delta_{il}\delta_{kj}-{1\over 2N_c}\delta_{ij}\delta_{kl},
\eeq
and $PT$-invariance of the unpolarized state
\beq
\la P|\bar{\psi}(\lambda n)[\lambda n,0]gD_{\perp\alpha}(0)F^{\alpha n}(0)[0,\lambda n]
\slash{n}\psi(\lambda n)|P\ra=
\la P|\bar{\psi}(-\lambda n)[-\lambda n,0]gD_{\perp\alpha}(0)F^{\alpha n}(0)[0,-\lambda n]
\slash{n}\psi(-\lambda n)|P\ra.
\eeq
We calculate the last term in (\ref{gluon_condensate}),
\beq
T_3=4\la P|F_a^{in}(0)F_a^{pi}(0)|P\ra=
\int dx\Bigl[-{4\over x^2}\int{d\lambda\over 2\pi}e^{i\lambda x}
\la P|F^{in}(0)[0,\lambda n]D^n(\lambda n)D^n(\lambda n)F^{pi}(\lambda n)|P\ra\Bigr].
\eeq
Combining the Bianchi identity (\ref{Bianchi}) and the equations of motion (\ref{EOM}),
we can derive another identity
\beq
2D_{ab}^n(\lambda n)F_{b}^{pi}(\lambda n)=D_{ab}^i(\lambda n)F_{b}^{pn}(\lambda n)
-D_{\perp ab\,j}(\lambda n)F_b^{ji}(\lambda n)
-g\bar{\psi}(\lambda n)t^a\gamma^i\psi(\lambda n).
\eeq
Exchanging the covariant derivatives by using 
\beq
[D^{\mu}(\lambda n),D^{\nu}(\lambda n)]=-igF^{\mu\nu}(\lambda n),
\eeq
we obtain
\beq
T_3&=&-2\int{dx\over x^2}\int{d\lambda\over 2\pi}e^{i\lambda x}\Bigl\{
2\la P|F^{in}(0)[0,\lambda n]D_{\perp}^{j}(\lambda n)D_{\perp}^{i}(\lambda n)
F^{jn}(\lambda n)|P\ra
\nonumber\\
&&-\la P|F^{in}(0)[0,\lambda n]D^j_{\perp}(\lambda n)D_{\perp}^{j}(\lambda n)
F^{in}(\lambda n)|P\ra+i\la P|F^{in}(0)[0,\lambda n]gF^{in}(\lambda n)F^{pn}(\lambda n)|P\ra
\nonumber\\
&&+\la P|\bar{\psi}(0)\{0,\lambda n\}
D_{\perp}^i(\lambda n)\Bigl(gF^{in}(\lambda n)\{\lambda n,0\}\slash{n}\psi(0)\Bigr)|P\ra
\nonumber\\
&&+i\int^0_{\lambda}dt\,\la P|\bar{\psi}(0)\{0,tn\}gF^{in}(tn)
\{tn,\lambda n\}gF^{in}(\lambda n)\{\lambda n,0\}\slash{n}\psi(0)|P\ra
\nonumber\\
&&-\la P|\bar{\psi}(0)\{0,\lambda n\}gF^{in}(\lambda n)\{\lambda n,0\}\slash{n}
D_{\perp}^i(0)\psi(0)|P\ra
-ix\la P|\bar{\psi}(0)\{0,\lambda n\}
gF^{in}(\lambda n)\{\lambda n,0\}\gamma^i\psi(0)|P\ra\Bigr\},
\label{T_3}
\eeq
where we used
\beq
F_{ba}^{jn}F_a^{ji}=-{i\over g}[D^j,D^i]_{bc}F_c^{jn}.
\eeq
Combining (\ref{T_1}), (\ref{T_2}) and (\ref{T_3}),
we obtain the result for the gluon condensate term written 
in terms of lightcone matrix elements as
\beq
&&\la P|F_a^{\mu\nu}(0)F_a^{\mu\nu}(0)|P\ra
\nonumber\\
&=&T_1+T_2+T_3
\nonumber\\
&=&-2\int {dx\over x^2}\Bigl\{
(g_{\perp\mu\nu}g_{\perp\alpha\beta}-g_{\perp\mu\beta}g_{\perp\alpha\nu}
-g_{\perp\mu\alpha}g_{\perp\nu\beta})M^{G\,\mu\nu\alpha\beta}_1(x)
+(2g_{\perp\mu\beta}g_{\perp\nu\alpha}-g_{\perp\mu\nu}g_{\perp\alpha\beta})
M^{G\,\mu\nu\alpha\beta}_2(x)+iM^{G}_{3}(x)
\nonumber\\
&&-ixM^{Q}_{1}(x)+2M^{Q}_{2}(x)+M^{Q}_{3}(x)-M^{Q}_{4}(x)
-{1\over 2}(\slash{n})_{ji}(\slash{n})_{lk}M^{4Q}_{1\,ijkl}(x)
-{1\over 2N_c}(\slash{n})_{li}(\slash{n})_{jk}M^{4Q}_{2\,ijkl}(x)
\Bigr\},
\label{gluon_condensate_expansion}
\eeq
where each matrix element is given by
\beq
M^{G\,\mu\nu\alpha\beta}_1(x)&=&\int{d\lambda\over 2\pi}e^{i\lambda x}\Bigl[
\la P|F^{\nu n}(0)[0,\lambda n]D_{\perp}^{\beta}(\lambda n)
D_{\perp}^{\alpha}(\lambda n)F^{\mu n}(\lambda n)|P\ra
\nonumber\\
&&+i\int^{0}_{\lambda}
d\mu\,\la P|F^{\nu n}(0)[0,\mu n]gF^{\beta n}(\mu n)[\mu n,\lambda n]
D_{\perp}^{\alpha}(\lambda n)F^{\mu n}(\lambda n)|P\ra
\Bigr],
\eeq
\beq
M^{G\,\mu\nu\alpha\beta}_2(x)&=&\int{d\lambda\over 2\pi}e^{i\lambda x}\la P|F^{\nu n}(0)
[0,\lambda n]D_{\perp}^{\beta}(\lambda n)D_{\perp}^{\alpha}(\lambda n)F^{\mu n}(\lambda n)|P\ra,
\eeq
\beq
M^{G}_{3}(x)=\int{d\lambda\over 2\pi}e^{i\lambda x}
\la P|F^{in}(0)[0,\lambda n]gF^{in}(\lambda n)F^{pn}(\lambda n)|P\ra,
\eeq
\beq
M^{Q}_{1}(x)=\int{d\lambda\over 2\pi}e^{i\lambda x}
\la P|\bar{\psi}(0)\{0,\lambda n\}
gF^{in}(\lambda n)\{\lambda n,0\}\gamma^i\psi(0)|P\ra,
\eeq
\beq
M^{Q}_{2}(x)=\int{d\lambda\over 2\pi}e^{i\lambda x}
\la P|\bar{\psi}(0)\{0,\lambda n\}g\Bigl(D_{\perp i}(\lambda n)F^{i n}(\lambda n)\Bigr)
\{\lambda n,0\}\slash{n}\psi(0)|P\ra,
\eeq
\beq
M^{Q}_{3}(x)&=&\int{d\lambda\over 2\pi}e^{i\lambda x}\Bigl[\la P|\bar{\psi}(0)\{0,\lambda n\}
D_{\perp}^i(\lambda n)\Bigl(gF^{in}(\lambda n)\{\lambda n,0\}\slash{n}\psi(0)\Bigr)|P\ra
\nonumber\\
&&+i\int^0_{\lambda}dt\,\la P|\bar{\psi}(0)\{0,tn\}gF^{in}(tn)
\{tn,\lambda n\}gF^{in}(\lambda n)\{\lambda n,0\}\slash{n}\psi(0)|P\ra\Bigr],
\eeq
\beq
M^{Q}_{4}(x)&=&\int{d\lambda\over 2\pi}e^{i\lambda x}
\la P|\bar{\psi}(0)\{0,\lambda n\}gF^{in}(\lambda n)\{\lambda n,0\}\slash{n}
D_{\perp}^i(0)\psi(0)|P\ra,
\eeq
\beq
M^{4Q}_{1\,ijkl}(x)&=&g^2\int{d\lambda\over 2\pi}e^{i\lambda x}
\la P|\bar{\psi}_j(0)\{0,\lambda n\}\psi_k(\lambda n)
\bar{\psi}_l(\lambda n)\{\lambda n,0\}\psi_i(0)|P\ra,
\eeq
\beq
M^{4Q}_{2\,ijkl}(x)&=&g^2\int{d\lambda\over 2\pi}e^{i\lambda x}
\la P|\bar{\psi}_j(0)\psi_k(0)
\bar{\psi}_l(\lambda n)\psi_i(\lambda n)|P\ra.
\eeq
Our goal in this section is to rewrite each matrix element in terms of experimentally 
accessible twist-4 matrix elements.

\

\noindent
\underline{$M^{G\,\mu\nu\alpha\beta}_1(x)$, $M^{G\,\mu\nu\alpha\beta}_2(x)$}

\

We introduce the $DD$-type matrix element
\beq
&&M^{\mu\nu\alpha\beta}_{DD}(x_1,x_2,x_3)
\nonumber\\
&=&\int{d\lambda\over 2\pi}\,e^{i\lambda x_1}\int{d\mu\over 2\pi}\,e^{i\mu (x_3-x_1)}
\int{d\omega\over 2\pi}\,e^{i\omega (x_2-x_3)}
\la P|F^{\nu n}(0)[0,\omega n]D_{\perp}^{\beta}(\omega n)[\omega n,\mu n]
D_{\perp}^{\alpha}(\mu n)[\mu n,\lambda n]F^{\mu n}(\lambda n)|P\ra.
\eeq
Using the formula
\beq
D^{\alpha}_{\perp}(\mu n)[\mu n,\lambda n]=[\mu n,\lambda n]D^{\alpha}_{\perp}(\lambda n)
+i\int^{\mu}_{\lambda}dt\,[\mu n,t n]gF^{\alpha n}(tn)[t n,\lambda n],
\eeq
The $DD$-type matrix element can be rewritten as
\beq
&&M^{\mu\nu\alpha\beta}_{DD}(x_1,x_2,x_3)
\nonumber\\
&&=\delta(x_3-x_1)\delta(x_2-x_3)
\int{d\lambda\over 2\pi}\,e^{i\lambda x_1}
\la P|F^{\nu n}(0)[0,\lambda n]D_{\perp}^{\beta}(\lambda n)
D_{\perp}^{\alpha}(\lambda n)F^{\mu n}(\lambda n)|P\ra
\nonumber\\
&&+i\delta(x_3-x_1)
\int{d\lambda\over 2\pi}\,e^{i\lambda x_1}\int{d\omega\over 2\pi}\,e^{i\omega (x_2-x_3)}
\int^{\omega}_{\lambda}dt\,
\la P|F^{\nu n}(0)[0,t n]gF^{\beta n}(t n)[t n,\lambda n]
D_{\perp}^{\alpha}(\lambda n)F^{\mu n}(\lambda n)|P\ra
\nonumber\\
&&+i\int{d\lambda\over 2\pi}\int{d\mu\over 2\pi}\int{d\omega\over 2\pi}
\,e^{i\lambda x_1}e^{i\mu (x_3-x_1)}
e^{i\omega (x_2-x_3)}\int^{\mu}_{\lambda} dt
\la P|F^{\nu n}(0)[0,\omega n]D_{\perp}^{\beta}(\omega n)[\omega n,t n]
gF^{\alpha n}(t n)[t n,\lambda n]F^{\mu n}(\lambda n)|P\ra.
\nonumber\\
\eeq
We find the relations
\beq
\int dx_1\int dx_2\,M^{\mu\nu\alpha\beta}_{DD}(x_1,x_2,x)
&=&M^{G\,\mu\nu\alpha\beta}_1(x),
\label{DD1}\\
\int dx_1\int dx_3\,M^{\mu\nu\alpha\beta}_{DD}(x_1,x,x_3)
&=&M^{G\,\mu\nu\alpha\beta}_2(x).
\label{DD2}
\eeq
On the other hand, $M^{\mu\nu\alpha\beta}_{DD}(x_1,x_2,x_3)$
can be written in terms of other matrix elements by
splitting the integral regions as follows:
\beq
\int^{\omega}_{\lambda}dt=\int dt\Bigl(\theta(t-\lambda)-\theta(t-\omega)\Bigr)
=\int^{\infty}_{\lambda} dt
-{1\over 2\pi i}\int dt\int dy{1\over y-i\epsilon}e^{i(t-\omega)y}.
\eeq
We can show
\beq
&&M^{\mu\nu\alpha\beta}_{DD}(x_1,x_2,x_3)
\nonumber\\
&=&\delta(x_3-x_1)\delta(x_2-x_3)M^{G\,\mu\nu\alpha\beta}_{\partial\partial}(x_3)
-\delta(x_3-x_1){1\over x_2-x_3-i\epsilon}
M^{G\,\mu\nu\beta\alpha}_{1\partial}(x_3,x_2)
\nonumber\\
&&-\delta(x_2-x_3){1\over x_3-x_1-i\epsilon}M^{G\,\mu\nu\alpha\beta}_{2\partial}(x_1,x_3)
+{1\over x_3-x_1-i\epsilon}{1\over x_2-x_3-i\epsilon}
M^{G\,\mu\nu\alpha\beta}_{FF}(x_1,x_2,x_3),
\label{DD3}
\eeq
where the matrix elements on the r.h.s are given by
\beq
M^{G\,\mu\nu\alpha\beta}_{\partial\partial}(x)&=&
\int{d\lambda\over 2\pi}\,e^{i\lambda x}
\la P|F^{\nu n}(0)[0,\lambda n]D_{\perp}^{\beta}(\lambda n)
D_{\perp}^{\alpha}(\lambda n)F^{\mu n}(\lambda n)|P\ra
\nonumber\\
&&+i\int{d\lambda\over 2\pi}\,e^{i\lambda x}
\int^{\infty}_{\lambda}dt\,
\la P|F^{\nu n}(0)[0,t n]gF^{\beta n}(t n)[t n,\lambda n]
D_{\perp}^{\alpha}(\lambda n)F^{\mu n}(\lambda n)|P\ra
\nonumber\\
&&+i\int{d\lambda\over 2\pi}\,e^{i\lambda x}
\int^{\infty}_{\lambda} dt\,
\la P|F^{\nu n}(0)[0,t n]D_{\perp}^{\beta}(t n)\Bigl(
gF^{\alpha n}(t n)[t n,\lambda n]F^{\mu n}(\lambda n)\Bigr)|P\ra
\nonumber\\
&&-\int{d\lambda\over 2\pi}\,e^{i\lambda x}
\int^{\infty}_{\lambda} dt\int^{\infty}_{t}dt'\,
\la P|F^{\nu n}(0)[0,t' n]gF^{\beta n}(t' n)[t' n,t n]
gF^{\alpha n}(t n)[t n,\lambda n]F^{\mu n}(\lambda n)|P\ra,
\label{gluon_doublekinematical}
\eeq
\beq
&&M^{G\,\mu\nu\alpha\beta}_{1\partial}(x_1,x_2)
\nonumber\\
&=&\int{d\lambda\over 2\pi}\,e^{i\lambda x_1}
\int {dt\over 2\pi}e^{it(x_2-x_1)}
\la P|F^{\nu n}(0)[0,t n]gF^{\alpha n}(t n)[t n,\lambda n]
D_{\perp}^{\beta}(\lambda n)F^{\mu n}(\lambda n)|P\ra
\nonumber\\
&&+i\int{d\lambda\over 2\pi}\,e^{i\lambda x_1}
\int{dt\over 2\pi}e^{it(x_2-x_1)}
\int^{\infty}_{\lambda} dt'\la P|F^{\nu n}(0)[0,t n]gF^{\alpha n}(t n)
[t n,t' n]gF^{\beta n}(t' n)[t' n,\lambda n]F^{\mu n}(\lambda n)|P\ra,
\nonumber\\
\label{gluon_kinematical_1}
\eeq
\beq
&&M^{G\,\mu\nu\alpha\beta}_{2\partial}(x_1,x_2)
\nonumber\\
&=&\int{d\lambda\over 2\pi}\,e^{i\lambda x_1}
\int {dt\over 2\pi}e^{it(x_2-x_1)}
\la P|F^{\nu n}(0)[0,t n]D_{\perp}^{\beta}(t n)\Bigl[
gF^{\alpha n}(t n)[t n,\lambda n]F^{\mu n}(\lambda n)\Bigr]|P\ra
\nonumber\\
&&+i\int{d\lambda\over 2\pi}\,e^{i\lambda x_1}
\int {dt\over 2\pi}e^{it(x_2-x_1)}\int^{\infty}_tdt'\,
\la P|F^{\nu n}(0)[0,t' n]gF^{\beta n}(t' n)[t' n,t n]
gF^{\alpha n}(t n)[t n,\lambda n]F^{\mu n}(\lambda n)|P\ra,
\nonumber\\
\label{gluon_kinematical_2}
\eeq
\beq
M^{G\,\mu\nu\alpha\beta}_{FF}(x_1,x_2,x_3)
&=&
\int{d\lambda\over 2\pi}\,e^{i\lambda x_1}
\int {d\mu\over 2\pi}e^{i\mu(x_3-x_1)}\int{d\omega\over 2\pi}e^{i\omega (x_2-x_3)}\,
\nonumber\\
&&\times\la P|F^{\nu n}(0)[0,\omega n]gF^{\beta n}(\omega n)[\omega n,\mu n]
gF^{\alpha n}(\mu n)[\mu n,\lambda n]F^{\mu n}(\lambda n)|P\ra.
\label{gluon_FF}
\eeq
We find that the matrix elements including a covariant derivative $D_{\perp}$
can be expressed as $k_T$-moments of TMD matrix elements,
\beq
M^{G\,\mu\nu\alpha\beta}_{\partial\partial}(x)
&=&-\int d^2k_T\,k_T^{\alpha}k_T^{\beta}M^{G\,\mu\nu}(x,k_T),
\nonumber\\
M^{G\,\mu\nu\alpha\beta}_{1\partial}(x_1,x_2)&=&-i
\int d^2k_{1T}\int d^2k_{2T}\,k^{\beta}_{1T}M^{G\,\mu\nu\alpha}(x_1,x_2,k_{1T},k_{2T}),
\nonumber\\
M^{G\,\mu\nu\alpha\beta}_{2\partial}(x_1,x_2)&=&-i
\int d^2k_{1T}\int d^2k_{2T}\,k^{\beta}_{2T}M^{G\,\mu\nu\alpha}(x_1,x_2,k_{1T},k_{2T}),
\eeq
where
\beq
&&M^{G\,\mu\nu}(x,k_T)
\nonumber\\
&=&\int{d\lambda\over 2\pi}e^{i\lambda x}\int{d\xi_{\perp}\over (2\pi)^2}
e^{ik_{T}\cdot \xi_{\perp}}
\la P|F^{\nu n}(0)[0,\infty n][\infty n,\infty n+\xi_{\perp}]
[\infty n+\xi_{\perp},\lambda n+\xi_{\perp}]F^{\mu n}(\lambda n+\xi_{\perp})|P\ra,
\eeq
\beq
M^{G\,\mu\nu\alpha}(x_1,x_2,k_{1T},k_{2T})
&=&\int{d\lambda\over 2\pi}e^{i\lambda x_1}\int{d\mu\over 2\pi}e^{i\mu (x_2-x_1)}
\int{d\xi_{\perp}\over (2\pi)^2}e^{ik_{1T}\cdot \xi_{\perp}}
\int{d\eta_{\perp}\over (2\pi)^2}e^{i(k_{2T}-k_{1T})\cdot \eta_{\perp}}
\la P|F^{\nu n}(0)[0,\infty n]
\nonumber\\
&&\times[\infty n,\infty n+\eta_{\perp}][\infty n+\eta_{\perp},\mu n+\eta_{\perp}]
gF^{\alpha n}(\mu n+\eta_{\perp})
[\mu n+\eta_{\perp},\infty n+\eta_{\perp}]
\nonumber\\
&&\times[\infty n+\eta_{\perp},\infty n+\xi_{\perp}]
[\infty n+\xi_{\perp},\lambda n+\xi_{\perp}]
F^{\mu n}(\lambda n+\xi_{\perp})|P\ra.
\eeq
These relations are important because they may allow us to partially evaluate 
the proton mass in terms of lower-twist TMD functions.
Substituting (\ref{DD3}) into (\ref{DD1}) and (\ref{DD2}), we obtain
\beq
M^{G\,\mu\nu\alpha\beta}_1(x)&=&
M^{G\,\mu\nu\alpha\beta}_{\partial\partial}(x)
-\int dx'\,{1\over x'-x-i\epsilon}
M^{G\,\mu\nu\beta\alpha}_{1\partial}(x,x')
-\int dx'\,{1\over x-x'-i\epsilon}M^{G\,\mu\nu\alpha\beta}_{2\partial}(x',x)
\nonumber\\
&&+\int dx'\int dx''\,{1\over x-x-i\epsilon}{1\over x''-x-i\epsilon}
M^{G\,\mu\nu\alpha\beta}_{FF}(x',x'',x),
\\
M^{G\,\mu\nu\alpha\beta}_2(x)&=&
M^{G\,\mu\nu\alpha\beta}_{\partial\partial}(x)
-\int dx'\,{1\over x-x'-i\epsilon}
M^{G\,\mu\nu\beta\alpha}_{1\partial}(x',x)
-\int dx'\,{1\over x-x'-i\epsilon}M^{G\,\mu\nu\alpha\beta}_{2\partial}(x',x)
\nonumber\\
&&+\int dx'\int dx''{1\over x''-x'-i\epsilon}{1\over x-x''-i\epsilon}
M^{G\,\mu\nu\alpha\beta}_{FF}(x',x,x'').
\eeq

\

\noindent
\underline{$M^{G}_{3}(x)$}

\

We define the matrix element
\beq
M^G_F(x_1,x_2)&=&\int{d\lambda\over 2\pi}\int{d\mu\over 2\pi}\,e^{i\lambda x_1}e^{i\mu(x_2-x_1)}
\la P|F^{in}(0)[0,\mu n]gF^{in}(\mu n)[\mu n,\lambda n]F^{pn}(\lambda n)|P\ra,
\eeq
which satisfies
\beq
M^{G}_{3}(x)=\int dx_1\,M^G_F(x_1,x).
\eeq
We use the equations of motion (\ref{EOM}) for $F^{pn}(\lambda n)$,
\beq
\nonumber\\
M^G_F(x_1,x_2)
&=&{i\over x_1}\int{d\lambda\over 2\pi}\int{d\mu\over 2\pi}\,e^{i\lambda x_1}e^{i\mu (x_2-x_1)}
\la P|F^{in}(0)[0,\mu n]gF^{in}(\mu n)[\mu n,\lambda n]D^n(\lambda n)F^{pn}(\lambda n)|P\ra
\nonumber\\
&=&{i\over x_1}\int{d\lambda\over 2\pi}\int{d\mu\over 2\pi}\,e^{i\lambda x_1}e^{i\mu (x_2-x_1)}
\la P|F^{in}(0)[0,\mu n]gF^{in}(\mu n)[\mu n,\lambda n]D_{\perp}^j(\lambda n)F^{jn}(\lambda n)|P\ra
\nonumber\\
&&-{i\over x_1}\int{d\lambda\over 2\pi}\int{d\mu\over 2\pi}\,e^{i\lambda x_2}e^{i\mu (x_1-x_2)}
\la P|\bar{\psi}(0)\{0,\mu n\}gF^{in}(\mu n)\{\mu n,\lambda n\}gF^{in}(\lambda n)\{\lambda n,0\}
\slash{n}\psi(0)|P\ra
\nonumber\\
&&+{i\over x_1}\int{d\lambda\over 2\pi}\int{d\mu\over 2\pi}\,e^{i\lambda x_2}e^{i\mu (x_1-x_2)}
\la P|\bar{\psi}(0)\{0,\lambda n\}gF^{in}(\lambda n)\{\lambda n,\mu n\}gF^{in}(\mu n)\{\mu n,0\}
\slash{n}\psi(0)|P\ra
\nonumber\\
&=&{i\over x_1}g_{\perp\mu\beta}g_{\perp\nu\alpha}M^{\prime \mu\nu\alpha\beta}(x_1,x_2)
+{i\over x_1}g_{\perp\alpha\beta}\int dx'\,{\rm Tr}[\slash{n}M^{Q\,\alpha\beta}_{FF}(x',x_1+x',x_2+x')]
\nonumber\\
&&-{i\over x_1}g_{\perp\alpha\beta}\int dx'\,{\rm Tr}[\slash{n}M^{Q\,\alpha\beta}_{FF}(x',x_1+x',x_1-x_2+x')],
\eeq
where we have defined twist-4 quark-gluon matrix element
\beq
&&M^{Q\,\alpha\beta}_{FF\,ij}(x_1,x_2,x_3)
\nonumber\\
&=&\int{d\lambda\over 2\pi}\int{d\mu\over 2\pi}\int{d\omega\over 2\pi}
\,e^{i\lambda x_1}e^{i\mu (x_3-x_1)}e^{i\omega (x_2-x_3)}
\la P|\bar{\psi}_j(0)\{0,\omega n\}gF^{\beta n}(\omega n)\{\omega n,\mu n\}gF^{\alpha n}(\mu n)
\{\mu n,\lambda n\}\psi_i(\lambda n)|P\ra.\hspace{5mm}
\label{quark_FF}
\eeq
For the calculation of the gluon part $M^{\prime \mu\nu\alpha\beta}(x_1,x_2)$,
we introduce a $FD$ type matrix element
\beq
&&M^{\mu\nu\alpha\beta}_{FD}(x_1,x_2,x_3)
\nonumber\\
&=&\int{d\lambda\over 2\pi}\int{d\mu\over 2\pi}\int{d\omega\over 2\pi}
\,e^{i\lambda x_1}e^{i\mu (x_3-x_1)}e^{i\omega (x_2-x_3)}
\la P|F^{\nu n}(0)[0,\omega n]gF^{\alpha n}(\omega n)[\omega n,\mu n]D_{\perp}^{\beta}(\mu n)
[\mu n,\lambda n]F^{\mu n}(\lambda n)|P\ra.\hspace{3mm}
\eeq
Using the same technique as for $DD$ type matrix element, we can show the relations
\beq
\int dx'\,M^{\mu\nu\alpha\beta}_{FD}(x',x_2,x_1)
&=&M^{\prime \mu\nu\alpha\beta}(x_1,x_2),
\eeq
and
\beq
M^{\mu\nu\alpha\beta}_{FD}(x_1,x_2,x_3)&=
&\delta(x_3-x_1)M^{G\,\mu\nu\alpha\beta}_{1\partial}(x_1,x_2)
-{1\over x_3-x_1-i\epsilon}M^{G\,\mu\nu\beta\alpha}_{FF}(x_1,x_2,x_3).
\eeq 
We finally express $M^{G}_{3}(x)$ in terms of the twist-4 matrix elements,
\beq
M^{G}_{3}(x)&=&\int dx'\,M^G_F(x',x)
\nonumber\\
&=&i\int {dx'\over x'}\Bigl\{g_{\perp\mu\beta}g_{\perp\nu\alpha}\Bigl[
M^{G\,\mu\nu\alpha\beta}_{1\partial}(x',x)
-\int dx''{1\over x'-x''-i\epsilon}M^{G\,\mu\nu\beta\alpha}_{FF}(x'',x,x')\Bigr]
\nonumber\\
&&+g_{\perp\alpha\beta}\int dx''\,{\rm Tr}[\slash{n}M^{Q\,\alpha\beta}_{FF}(x'',x'+x'',x+x'')]
-g_{\perp\alpha\beta}\int dx''\,{\rm Tr}[\slash{n}M^{Q\,\alpha\beta}_{FF}(x'',x'+x'',x'-x+x'')]\Bigr\}.
\hspace{5mm}
\eeq

\

\noindent
\underline{$M^{Q}_{1}(x)$}

\

We introduce the matrix element
\beq
&&M^Q_F(x_1,x_2)=\int{d\lambda\over 2\pi}\int{d\mu\over 2\pi}e^{i\lambda x_1}e^{i\mu(x_2-x_1)}
\la P|\bar{\psi}(0)\{0,\mu n\}gF^{in}(\mu n)\{\mu n,\lambda n\}\gamma^i\psi(\lambda n)|P\ra,
\eeq
which satisfies
\beq
M^{Q}_{1}(x)=\int dx_1\,M^Q_F(x_1,x+x_1).
\eeq
The quark bilinear form $\bar{\psi}\gamma_i\psi$ contains a ``bad component'' of the quark field,
and it is written solely in terms of the good component using the equations of
motion\cite{Jaffe:1996zw,Kanazawa:2015ajw}.
We insert $1=\{\slash{p},\slash{n}\}/2$ in the matrix element,
\beq
M^Q_F(x_1,x_2)&=&{1\over 2}\int{d\lambda\over 2\pi}\int{d\mu\over 2\pi}e^{i\lambda x_1}
e^{i\mu(x_2-x_1)}
\la P|\bar{\psi}(0)\{0,\mu n\}gF^{in}(\mu n)\{\mu n,\lambda n\}
\gamma^i(\slash{n}\slash{p}+\slash{p}\slash{n})\psi(\lambda n)|P\ra
\nonumber\\
&=&{i\over 2x_1}\int{d\lambda\over 2\pi}\int{d\mu\over 2\pi}
e^{i\lambda x_1}e^{i\mu(x_2-x_1)}
\la P|\bar{\psi}(0)\{0,\mu n\}gF^{in}(\mu n)\{\mu n,\lambda n\}\gamma^i\slash{n}
\slash{p}D_{\perp}^n(\lambda n)\psi(\lambda n)|P\ra
\nonumber\\
&&-{i\over 2x_2}\int{d\lambda\over 2\pi}\int{d\mu\over 2\pi}e^{i\lambda x_1}e^{i\mu(x_2-x_1)}
\la P|\bar{\psi}(0)\overleftarrow{D}^n(0)\slash{p}\slash{n}\gamma^i\{0,\mu n\}
gF^{in}(\mu n)\{\mu n,\lambda n\}
\psi(\lambda n)|P\ra.
\eeq
Using the Dirac equation
\beq
\slash{D}\psi=\slash{p}D^n+\slash{n}D^p+\gamma_jD_{\perp}^j=0
\hspace{5mm}\to\hspace{5mm}
\slash{n}\slash{p}D^n+\slash{n}\gamma_jD_{\perp}^j=0,
\eeq
we can show
\beq
&&M^Q_F(x_1,x_2)
\nonumber\\
&=&-{i\over 2x_1}\int{d\lambda\over 2\pi}\int{d\mu\over 2\pi}
e^{i\lambda x_1}e^{i\mu(x_2-x_1)}
\la P|\bar{\psi}(0)\{0,\mu n\}gF^{in}(\mu n)\{\mu n,\lambda n\}\gamma^i\slash{n}
\gamma_jD_{\perp}^j(\lambda n)\psi(\lambda n)|P\ra
\nonumber\\
&&+{i\over 2x_2}\int{d\lambda\over 2\pi}\int{d\mu\over 2\pi}e^{i\lambda x_1}e^{i\mu(x_2-x_1)}
\la P|\bar{\psi}(0)\overleftarrow{D}^j(0)\gamma_j\slash{n}\gamma^i\{0,\mu n\}
gF^{in}(\mu n)\{\mu n,\lambda n\}
\psi(\lambda n)|P\ra
\nonumber\\
&=&
-{i\over 2x_1}\int{d\lambda\over 2\pi}\int{d\mu\over 2\pi}
e^{i\lambda x_1}e^{i\mu(x_2-x_1)}
\la P|\bar{\psi}(0)\{0,\mu n\}gF^{in}(\mu n)\{\mu n,\lambda n\}\gamma^i\slash{n}\gamma_j
D^j(\lambda n)\psi(\lambda n)|P\ra
\nonumber\\
&&-{i\over 2x_2}\int{d\lambda\over 2\pi}\int{d\mu\over 2\pi}e^{i\lambda x_1}e^{i\mu(x_2-x_1)}
\la P|\bar{\psi}(0)\gamma_j\slash{n}\gamma^i\{0,\mu n\}
D^j(\mu n)\Bigl(gF^{in}(\mu n)\{\mu n,\lambda n\}
\psi(\lambda n)\Bigr)|P\ra
\nonumber\\
&&+{1\over 2x_2}\int{d\lambda\over 2\pi}\int{d\mu\over 2\pi}e^{i\lambda x_1}e^{i\mu(x_2-x_1)}
\int^{0}_{\mu}dt\,
\la P|\bar{\psi}(0)\gamma_j\slash{n}\gamma^i\{0,t n\}
gF^{jn}(t n)
\{tn,\mu n\}gF^{in}(\mu n)\{\mu n,\lambda n\}
\psi(\lambda n)|P\ra
\nonumber\\
&=&-{i\over 2x_1}\int dx''\,{\rm Tr}[\gamma^i\slash{n}\gamma_jM^{Q\,ij}_{FD}(x'',x_2,x_1)]
-{i\over 2x_2}\int dx''\,{\rm Tr}[\gamma_j\slash{n}\gamma^iM^{Q\,ij}_{DF}(x_1,x'',x_2)],
\eeq
where we have defined quark-gluon $FD$ and $DF$ type functions
\beq
&&M^{Q\,\alpha\beta}_{FD\,ij}(x_1,x_2,x_3)
\nonumber\\
&=&\int{d\lambda\over 2\pi}\,e^{i\lambda x_1}\int{d\mu\over 2\pi}\,e^{i\mu (x_3-x_1)}
\int{d\omega\over 2\pi}\,e^{i\omega (x_2-x_3)}
\la P|\bar{\psi}_j(0)\{0,\omega n\}gF^{\alpha n}(\omega n)\{\omega n,\mu n\}
D_{\perp}^{\beta}(\mu n)\{\mu n,\lambda n\}\psi_i(\lambda n)|P\ra
\nonumber\\
&=&\delta(x_3-x_1)M^{Q\,\alpha\beta}_{1\partial\,ij}(x_1,x_2)
-{1\over x_3-x_1-i\epsilon}M^{Q\,\beta\alpha}_{FF\,ij}(x_1,x_2,x_3),
\eeq
\beq
&&M^{Q\,\alpha\beta}_{DF\,ij}(x_1,x_2,x_3)
\nonumber\\
&=&\int{d\lambda\over 2\pi}\,e^{i\lambda x_1}\int{d\mu\over 2\pi}\,e^{i\mu (x_3-x_1)}
\int{d\omega\over 2\pi}\,e^{i\omega (x_2-x_3)}
\la P|\bar{\psi}_j(0)\{0,\omega n\}D_{\perp}^{\beta}(\omega n)\{\omega n,\mu n\}
gF^{\alpha n}(\mu n)\{\mu n,\lambda n\}\psi_i(\lambda n)|P\ra
\nonumber\\
&=&\delta(x_2-x_3)M^{Q\,\alpha\beta}_{2\partial\,ij}(x_1,x_2)
-{1\over x_2-x_3-i\epsilon}M^{Q\,\alpha\beta}_{FF\,ij}(x_1,x_2,x_3).
\eeq
$M^{Q\,\alpha\beta}_{1\partial\,ij}(x_1,x_2)$ and $M^{Q\,\alpha\beta}_{2\partial\,ij}(x_1,x_2)$
are TMD-related matrix elements given by
\beq
&&M^{Q\,\alpha\beta}_{1\partial\,ij}(x_1,x_2)
\nonumber\\
&=&\int{d\lambda\over 2\pi}\,e^{i\lambda x_1}
\int{d\omega\over 2\pi}\,e^{i\omega (x_2-x_1)}
\la P|\bar{\psi}_j(0)[0,\omega n]gF^{\alpha n}(\omega n)[\omega n,\lambda n]
D_{\perp}^{\beta}(\lambda n)\psi_i(\lambda n)|P\ra
\nonumber\\
&&+i\int{d\lambda\over 2\pi}\,e^{i\lambda x_1}
\int{d\omega\over 2\pi}\,e^{i\omega (x_2-x_1)}\int^{\infty}_{\lambda}dt\,
\la P|\bar{\psi}_j(0)[0,\omega n]gF^{\alpha n}(\omega n)[\omega n,t n]
gF^{\beta n}(t n)[t n,\lambda n]\psi_i(\lambda n)|P\ra,
\label{quark_kinematical1}
\eeq
\beq
&&M^{Q\,\alpha\beta}_{2\partial\,ij}(x_1,x_2)
\nonumber\\
&=&\int{d\lambda\over 2\pi}\,e^{i\lambda x_1}
\int{d\mu\over 2\pi}\,e^{i\mu (x_2-x_1)}\,
\la P|\bar{\psi}_j(0)[0,\mu n]D_{\perp}^{\beta}(\mu n)\Bigl(
gF^{\alpha n}(\mu n)[\mu n,\lambda n]\psi_i(\lambda n)\Bigr)|P\ra
\nonumber\\
&&+i\int{d\lambda\over 2\pi}\,e^{i\lambda x_1}\int{d\mu\over 2\pi}\,e^{i\mu (x_2-x_1)}
\int^{\infty}_{\mu}dt\,\la P|\bar{\psi}_j(0)[0,t n]gF^{\beta n}(t n)[t n,\mu n]
gF^{\alpha n}(\mu n)[\mu n,\lambda n]\psi_i(\lambda n)|P\ra.
\label{quark_kinematical2}
\eeq
$M^{Q}_{1}(x)$ is finally given by
\beq
&&M^{Q}_{1}(x)
\nonumber\\
&=&\int dx_1\,M_F(x_1,x+x_1)
\nonumber\\
&=&-{i\over 2}\int dx'\Bigl[
{1\over x'}{\rm Tr}[\gamma^i\slash{n}\gamma_jM^{Q\,ij}_{1\partial}(x',x+x')]
-{1\over x'}\int dx''\,{1\over x'-x''-i\epsilon}
{\rm Tr}[\gamma^i\slash{n}\gamma_jM^{Q\,ji}_{FF}(x'',x+x',x')]
\nonumber\\
&&+{1\over x+x'}{\rm Tr}[\gamma_j\slash{n}\gamma^iM^{Q\,ij}_{2\partial}(x',x+x')]
-{1\over x+x'}\int dx''\,{1\over x''-x-x'-i\epsilon}
{\rm Tr}[\gamma_j\slash{n}\gamma^iM^{Q\,ij}_{FF}(x',x'',x+x')]
\Bigr].
\eeq

\

\noindent
\underline{$M^{Q}_{2}(x)$}

\

We can show the relation
\beq
&&\int dx_1\int dx_3\,\Bigl(M^{Q\,\alpha\beta}_{DF\,ij}(x_1,x_1+x_2,x_3)
-M^{Q\,\alpha\beta}_{FD\,ij}(x_1,x_1+x_2,x_3)\Bigr)
\nonumber\\
&=&\int{d\mu\over 2\pi}\,e^{i\mu x_2}
\la P|\bar{\psi}(0)\{0,\mu n\}\Bigl(gD_{\perp}^{\beta}(\mu n)F^{\alpha n}(\mu n)\Bigr)
\{\mu n,0\}\psi(0)|P\ra.
\eeq
Then we obtain
\beq
M^{Q}_{2}(x)&=&\int dx'\Bigl[
g_{\perp\alpha\beta}{\rm Tr}[\slash{n}M^{Q\,\alpha\beta}_{2\partial}(x',x'+x)]
-g_{\perp\alpha\beta}{\rm Tr}[\slash{n}M^{Q\,\alpha\beta}_{1\partial}(x',x'+x)]
\nonumber\\
&&-\int dx''\,\Bigl({1\over x'+x-x''-i\epsilon}-{1\over x''-x'-i\epsilon}\Bigr)
g_{\perp\alpha\beta}{\rm Tr}[\slash{n}M^{Q\,\alpha\beta}_{FF}(x',x'+x,x'')]
\Bigr].
\eeq

\

\noindent
\underline{$M^{Q}_{3}(x)$}

\

Using the definition of $M^{\alpha\beta}_{DF\,ij}$, we can easily show
\beq
M^{Q}_{3}(x)&=&-g_{\perp\alpha\beta}\int dx'\int dx''\,{\rm Tr}[\slash{n}
M^{\alpha\beta}_{DF}(x',x'',x+x')]
\nonumber\\
&=&-g_{\perp\alpha\beta}\int dx'\Bigl[
{\rm Tr}[\slash{n}M^{Q\,\alpha\beta}_{2\partial}(x',x+x')]
-\int dx''\,{1\over x''-x-x'-i\epsilon}{\rm Tr}[\slash{n}M^{Q\,\alpha\beta}_{FF}(x',x'',x+x')]\Bigr].
\eeq

\

\noindent
\underline{$M^{Q}_{4}(x)$}

\

Similarly,
\beq
M^{Q}_{4}(x)&=&-g_{\perp\alpha\beta}\int dx'\int dx''\,
{\rm Tr}[\slash{n}M^{\alpha\beta}_{FD\,ij}(x'',x+x',x')]
\nonumber\\
&=&-g_{\perp\alpha\beta}\int dx'\Bigl[{\rm Tr}[\slash{n}M^{Q\,\alpha\beta}_{1\partial}(x',x+x')]
-\int dx''\,{1\over x'-x''-i\epsilon}{\rm Tr}[\slash{n}M^{Q\,\beta\alpha}_{FF}(x'',x+x',x')]\Bigr].
\eeq

\

\noindent
\underline{$M^{4Q}_{1\,ijkl}(x)$, $M^{4Q}_{2\,ijkl}(x)$}

\

We define the 4-quark matrix element
\beq
M^{4Q}_{ik:jl}(x_1,x_2,x_3)&=&g^2\int{d\lambda\over 2\pi}\int{d\mu\over 2\pi}\int{d\omega\over 2\pi}\,
e^{i\lambda x_1}e^{i\mu(x_3-x_1)}e^{i\omega(x_2-x_3)}
\nonumber\\
&&\times\la P|\bar{\psi}_{j}(0)\{0,\omega n\}\psi_{k}(\omega n)
\bar{\psi}_{l}(\mu n)\{\mu n,\lambda n\}\psi_{i}(\lambda n)|P\ra.
\label{4-quark}
\eeq
We can easily check the relations
\beq
M^{4Q}_{1\,ijkl}(x)&=&\int dx'\int dx''\,M^{4Q}_{ik:jl}(x',x+x',x''+x'),
\nonumber\\
M^{4Q}_{2\,ijkl}(x)&=&\int dx'\int dx''\,M^{4Q}_{ik:jl}(x',x'',x).
\eeq
We have shown that the matrix elements appearing in the gluon condensate term 
(\ref{gluon_condensate_expansion}) can be written in terms of the following set of 
matrix elements.
\beq
&&\Bigl\{M^{G\,\mu\nu\alpha\beta}_{\partial\partial}(x),\ 
M^{G\,\mu\nu\beta\alpha}_{1\partial}(x_1,x_2),\ 
M^{G\,\mu\nu\alpha\beta}_{2\partial}(x_1,x_2),\ 
M^{G\,\mu\nu\alpha\beta}_{FF}(x_1,x_2,x_3),\ 
M^{Q\,\alpha\beta}_{1\partial}(x_1,x_2),\ 
M^{Q\,\alpha\beta}_{2\partial}(x_1,x_2),\ 
\nonumber\\
&&M^{Q\,\alpha\beta}_{FF}(x_1,x_2,x_3),\ 
M^{4Q}_{ik:jl}(x_1,x_2,x_3)
\Bigr\}
\eeq
In the following sections, we will show that all these matrix elements appear in
twist-4 contributions to cross sections in high energy inclusive processes.
It is also important to point out that matrix elements including a $\partial$ subscript are related to 
TMD functions and may therefore be determined indirectly within the TMD framework.


\section{Derivation of the quark-gluon twist-4 contributions to the unpolarized cross section in $ep$ collisions}

\begin{figure}[H]
\begin{center}
  \includegraphics[height=4cm,width=16cm]{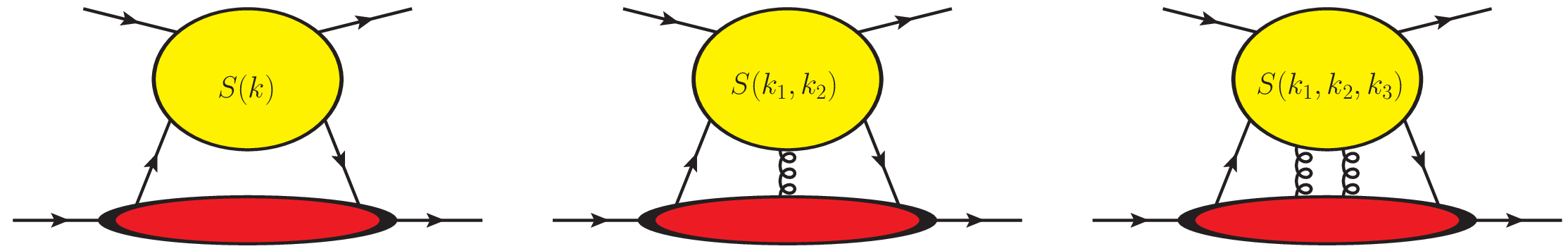}
\end{center}
 \caption{Series of diagrams considered in the diagrammatic method. Each diagram includes
 all twist contributions.}
\label{quark_diagrammatic}
\end{figure}

We discuss twist-4 contributions to the unpolarized cross section for light hadron production in
$ep$ collisions,
\beq
e(\ell)+p(P)\to h(P_h)+X.
\eeq
The unpolarized cross section is given by
\beq
P_h^0{d\sigma\over d^3\vec{P}_h}={\alpha_{em}^2\over S}\int{dz\over z^2}D(z)
W(P,\ell,{P_h\over z}),
\eeq
where $\alpha_{em}$ is the QED coupling, $S$ is the center of mass energy squared
and $D(z)$ is the fragmentation function of the final state hadron $h$.
$W$ describes the scattering of the electron and the proton. We derive the formula for twist-4 contributions 
to $W$ in accordance with the diagrammatic method\cite{Ellis:1982wd,Ellis:1982cd}.
We consider the three types of diagrams shown in Fig. \ref{quark_diagrammatic} at twist-4.
\noindent
The left diagram is given by
\beq
W^{\rm Fig.1left}=\int{d^4k\over (2\pi)^4}\int d^4\xi\,e^{ik\cdot \xi}\la P|\bar{\psi}_j(0)\psi_i(\xi)|P\ra 
S_{ji}(k).
\eeq
There is one diagram in $S_{ji}(k)$.
\begin{figure}[H]
\begin{center}
  \includegraphics[height=4cm,width=5cm]{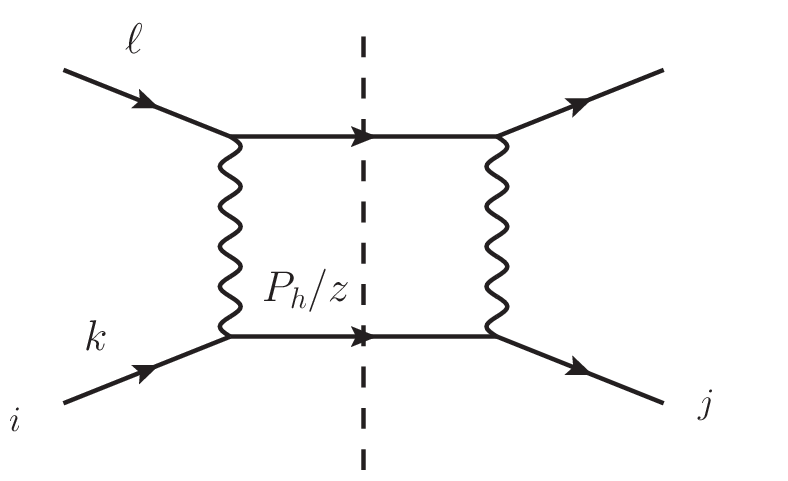}
\end{center}
 \caption{The hard scattering diagram in $S_{ji}(k)$.}
\end{figure}
\noindent
We should take into account the proton mass correction to the proton momentum $P$ at the twist-4 level,
\beq
P^{\mu}=p^{\mu}+{M_N^2\over 2}n^{\mu}.
\eeq
However, the parametrization of twist-4 distribution functions always contains a factor of $M_N^2$, 
which leads to a suppression of order $(M_N/Q)^2$ in the cross section. Therefore, proton mass corrections 
to the hard-scattering coefficients introduce an additional factor of $(M_N/Q)^2$ and contribute only at 
order $(M_N/Q)^4$ or higher. Thus, we can neglect these corrections in the twist-4 calculation and 
effectively regard $P$ as the lightlike vector $p$. We expand the momentum vector as
\beq
k^{\mu}=(k\cdot n)p^{\mu}+(k\cdot p)n^{\mu}+k_{\perp}^{\mu}.
\eeq
We perform the Taylor expansion around $k=(k\cdot n)p$ in order to extract the twist-4 contribution.
\beq
S(k)\simeq S((k\cdot n)p)+{\partial\over \partial k^{\mu}}S(k)\Bigr|_{k=(k\cdot n)p}
\omega^{\mu}_{\ \alpha}k^{\alpha}
+{1\over 2}{\partial^2\over \partial k^{\mu}\partial k^{\nu}}S(k)\Bigr|_{k=(k\cdot n)p}
\omega^{\mu}_{\ \alpha}k^{\alpha}\omega^{\nu}_{\ \beta}k^{\beta},
\label{Taylor1}
\eeq
where $\omega^{\mu}_{\alpha}=g^{\mu}_{\ \alpha}-p^{\mu}n_{\alpha}$.
The third term gives the twist-4 contribution. 
The vector $k$ is rewritten as
\beq
k^{\alpha}\hspace{5mm}\to\hspace{5mm}-i{\partial\over \partial \xi_{\alpha}}e^{ik\cdot \xi}.
\eeq
Integrating by parts with respect to $\xi$,
we obtain
\beq
W^{\rm Fig.1left}\Bigr|_{\rm twist-4}&=&{1\over 2}\omega^{\mu}_{\ \alpha}\omega^{\nu}_{\ \beta}
\int{d^4k\over (2\pi)^4}\int d^4\xi\,e^{ik\cdot \xi}\la P|\bar{\psi}_j(0)\psi_i(\xi)|P\ra
{\partial^2\over \partial k^{\mu}\partial k^{\nu}}S_{ji}(k)\Bigr|_{k=(k\cdot n)p}
k^{\alpha}k^{\beta}
\nonumber\\
&=&-{1\over 2}\omega^{\mu}_{\ \alpha}\omega^{\nu}_{\ \beta}
\int dx\int{d\lambda\over 2\pi}e^{i\lambda x}\la P|\bar{\psi}_j(0)
\partial^{\beta}\partial^{\alpha}\psi_i(\lambda n)|P\ra
{\partial^2\over \partial k^{\mu}\partial k^{\nu}}S_{ji}(k)\Bigr|_{k=xp},
\label{C_1_1}
\eeq
where we changed the integration variables to $k\cdot n=x$ and $\xi\cdot p=\lambda$.
We next consider the center diagram in Fig. \ref{quark_diagrammatic}. 
Its mathematical expression is given by
\beq
W^{\rm Fig.1center}=\int{d^4k_1\over (2\pi)^4}\int{d^4k_2\over (2\pi)^4}
\int d^4\xi\,e^{ik_1\cdot \xi}\int d^4\eta\,e^{i(k_2-k_1)\cdot \eta}
\la P|\bar{\psi}_{jj'}(0)gA_a^{\mu}(\eta)\psi_{ii'}(\xi)|P\ra S^a_{\mu\,ji:j'i'}(k_1,k_2).
\eeq
There are two diagrams in $S^a_{\mu\,ji:j'i'}(k_1,k_2)$.
\begin{figure}[H]
\begin{center}
  \includegraphics[height=4cm,width=9cm]{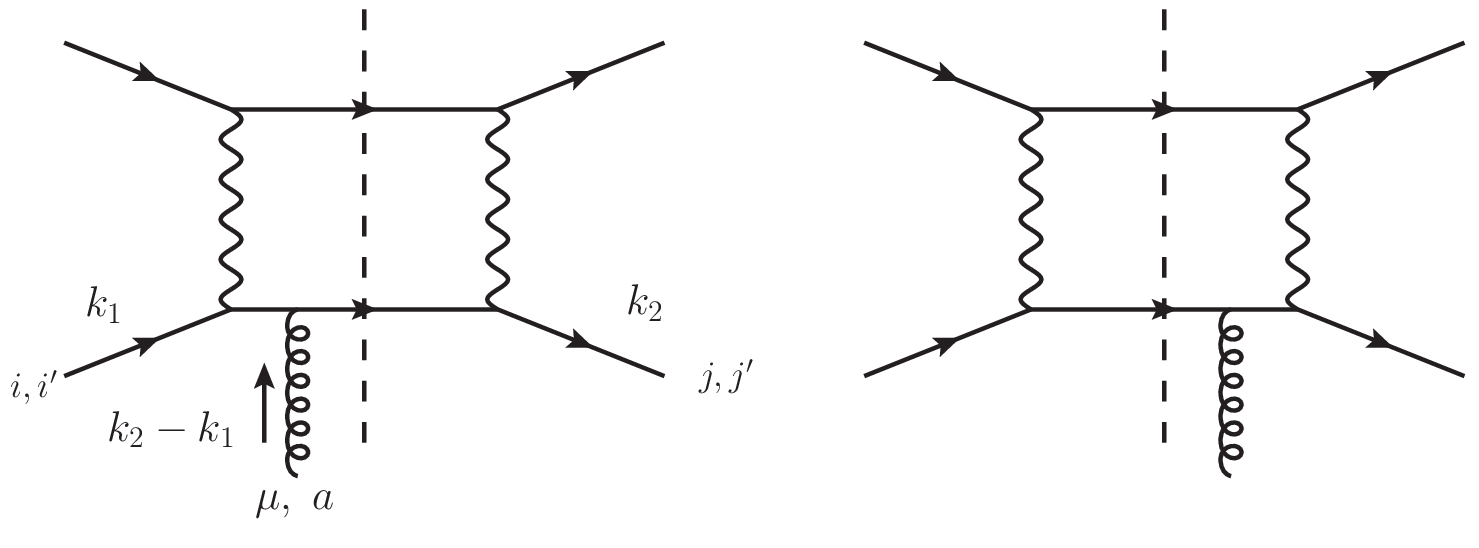}
\end{center}
 \caption{Hard scattering diagrams in $S^a_{\mu\,ji:j'i'}(k_1,k_2)$.}
\end{figure}
\noindent
We define $S_{\mu\,ji}(k_1,k_2)$ as
\beq
S^a_{\mu\,ji;j'i'}(k_1,k_2)=(t^a)_{j'i'}S_{\mu\,ji}(k_1,k_2)\hspace{5mm}
\to \hspace{5mm}S_{\mu\,ji}(k_1,k_2)={1\over N_cC_F}{\rm Tr}[t^aS^a_{\mu\,ji;j'i'}(k_1,k_2)].
\eeq
$S_{\mu\,ji}(k_1,k_2)$ satisfies the Ward-Takahashi identity(WTI)
\beq
(k_2-k_1)^{\mu}S_{\mu\,ji}(k_1,k_2)=S_{ji}(k_2)-S_{ji}(k_1).
\label{WTI1}
\eeq
We use the trick introduced in \cite{Beppu:2010qn}.
We rewrite the above WTI as
\beq
p^{\mu}S_{\mu\,ji}(k_1,k_2)={1\over (k_2-k_1)\cdot n-i\epsilon}
\Bigl[-\omega^{\mu}_{\ \alpha}(k_2-k_1)^{\alpha}S_{\mu\,ji}(k_1,k_2)
+\Bigl(S_{ji}(k_2)-S_{ji}(k_1)\Bigr)\Bigr],
\label{WTI_identity1}
\eeq
where the sign of $i\epsilon$ is determined by the fact that this scattering includes only 
the final-state interaction. In addition, we decompose the gluon field,
\beq
A_a^{\mu}=A^n_ap^{\mu}+\omega^{\mu}_{\ \alpha}A_a^{\alpha}.
\eeq
Then we can rewrite the gluon field in $W^{\rm Fig.1center}$,
\beq
A^{\mu}(\eta)S_{\mu\,ji}(k_1,k_2)&=&
-{1\over (k_2-k_1)\cdot n-i\epsilon}\omega^{\mu}_{\ \alpha}
\Bigl[A^n(\eta)(k_2-k_1)^{\alpha}
-[(k_2-k_1)\cdot n]A^{\alpha}(\eta)\Bigr]S_{\mu\,ji}(k_1,k_2)
\nonumber\\
&&+A^n(\eta){1\over (k_2-k_1)\cdot n-i\epsilon}
\Bigl(S_{ji}(k_2)-S_{ji}(k_1)\Bigr).
\eeq
Integrating by parts with respect to $\eta$, we can show
\beq
\Bigl[A^n(\eta)(k_2-k_1)^{\alpha}
-[(k_2-k_1)\cdot n]A^{\alpha}(\eta)\Bigr]\hspace{5mm}\to\hspace{5mm}
iF^{(0)\alpha n}(\eta),
\eeq
where $F^{(0)\alpha n}$ is the gluon's field strength tensor without the non-linear term,
$F^{(0)\alpha n}=\partial^{\alpha}A^n-\partial^nA^{\alpha}$.
Then we can show
\beq
&&W^{\rm Fig.1center}
\nonumber\\
&=&\int{d^4k_1\over (2\pi)^4}\int{d^4k_2\over (2\pi)^4}
\int d^4\xi\,e^{ik_1\cdot \xi}\int d^4\eta\,e^{i(k_2-k_1)\cdot \eta}
\Bigl[-i\omega^{\mu}_{\ \alpha}\la P|\bar{\psi}_{j}(0)gF^{(0)\alpha n}(\eta)\psi_{i}(\xi)|P\ra 
{1\over (k_2-k_1)\cdot n-i\epsilon}S_{\mu\,ji}(k_1,k_2)
\nonumber\\
&&+
\la P|\bar{\psi}_{j}(0)gA^n(\eta)\psi_{i}(\xi)|P\ra {1\over (k_2-k_1)\cdot n-i\epsilon}
\Bigr(S_{ji}(k_2)-S_{ji}(k_1)\Bigr)\Bigr].
\eeq
In order to extract twist-4 contributions(terms with 
$\omega^{\mu}_{\ \alpha}\omega^{\nu}_{\ \beta}$), we pick up the 
first derivative terms in the Taylor expansion of $S_{\mu\,ji}(k_1,k_2)$,
\beq
{\partial\over \partial k_1^{\nu}}S_{\mu\,ji}(k_1,k_2)\Bigr|_{k_i=(k_i\cdot n)p}
\omega^{\nu}_{\ \beta}k_1^{\beta}
+{\partial\over \partial k_2^{\nu}}S_{\mu\,ji}(k_1,k_2)\Bigr|_{k_i=(k_i\cdot n)p}
\omega^{\nu}_{\ \beta}k_2^{\beta}.
\eeq
After integration by parts, the first term in $W^{\rm Fig.1center}$ gives
\beq
&&\omega^{\mu}_{\ \alpha}\omega^{\nu}_{\ \beta}\int dx_1\int dx_2
\int{d\lambda\over 2\pi}\int{d\mu\over 2\pi}e^{i\lambda x_1}e^{i\mu (x_2-x_1)}
{1\over x_2-x_1-i\epsilon}
\nonumber\\
&&\times
\Bigl\{
\la P|\bar{\psi}_{j}(0)gF^{(0)\alpha n}(\mu n)\partial^{\beta}\psi_{i}(\lambda n)|P\ra 
{\partial\over \partial k_1^{\nu}}S_{\mu\,ji}(k_1,k_2)\Bigr|_{k_i=(k_i\cdot n)p}
\nonumber\\
&&+
\la P|\bar{\psi}_{j}(0)g\partial^{\beta}\Bigl(F^{(0)\alpha n}(\mu n)\psi_{i}(\lambda n)\Bigr)|P\ra 
{\partial\over \partial k_2^{\nu}}S_{\mu\,ji}(k_1,k_2)\Bigr|_{k_i=(k_i\cdot n)p}\Bigr\}.
\label{C_2_1}
\eeq
We pick up the second derivative term in (\ref{Taylor1}) for $S(k_{1,2})$.
The integration by parts with respect to $\eta$ is performed as
\beq
A^n(\eta)\psi_{i}(\xi)k^{\alpha}_2k^{\beta}_2\hspace{5mm}&\to&\hspace{5mm}
[iF^{(0)\alpha n}(\eta)\psi_{i}(\xi)+(k_2\cdot n-k_1\cdot n)A^{\alpha}(\eta)\psi_{i}(\xi)
+iA^n(\eta)\partial^{\alpha}\psi_{i}(\xi)]k^{\beta}_2
\nonumber\\
&\to&\hspace{5mm}
-\partial^{\beta}\Bigl(F^{(0)\alpha n}(\eta)\psi_{i}(\xi)\Bigr)
-(k_2\cdot n-k_1\cdot n)\partial^{\beta}\Bigl(-iA^{\alpha}(\eta)\psi_{i}(\xi)\Bigr)
\nonumber\\
&&\hspace{5mm}-F^{(0)\beta n}(\eta)\partial^{\alpha}\psi_{i}(\xi)
-(k_2\cdot n-k_1\cdot n)\Bigl(-iA^{\beta}(\eta)\Bigr)\partial^{\alpha}\psi_{i}(\xi)
-A^n(\eta)\partial^{\beta}\partial^{\alpha}\psi_{i}(\xi).
\eeq
Using the formula
\beq
\int dx_2{1\over x_2-x_1-i\epsilon}e^{i\mu x_2}=2\pi i\theta(\mu)e^{i\mu x_1},\hspace{5mm}
\int dx_1{1\over x_2-x_1-i\epsilon}e^{i(\lambda-\mu) x_1}
=2\pi i\theta(\mu-\lambda)e^{i(\lambda-\mu) x_2},
\eeq
The twist-4 contributions from the second term in $W^{\rm Fig.1center}$ are given by
\beq
&&-{1\over 2}\omega^{\mu}_{\ \alpha}\omega^{\nu}_{\ \beta}\int dx\int{d\lambda \over 2\pi}
e^{i\lambda x}
\Bigl\{
i\int^{\infty}_{\lambda}d\mu\,\la P|\bar{\psi}_{j}(0)g
\partial^{\beta}\Bigl[F^{(0)\alpha n}(\mu n)\psi_{i}(\lambda n)\Bigr]|P\ra 
\nonumber\\
&&+\la P|\bar{\psi}_{j}(0)\partial^{\beta}\Bigl[-igA^{\alpha}(\lambda n)\psi_{i}(\lambda n)\Bigr]|P\ra 
+i\int^{\infty}_{\lambda}d\mu\,\la P|\bar{\psi}_{j}(0)g
F^{(0)\beta n}(\mu n)\partial^{\alpha}\psi_{i}(\lambda n)|P\ra 
\nonumber\\
&&+\la P|\bar{\psi}_{j}(0)\Bigl(-igA^{\beta}(\lambda n)\Bigr)\partial^{\alpha}\psi_{i}(\lambda n)|P\ra 
+\la P|\bar{\psi}_{j}(0)\Bigl(ig\int^0_{\lambda}d\mu\,A^n(\mu n)\Bigr)
\partial^{\beta}\partial^{\alpha}\psi_{i}(\lambda n)|P\ra 
\Bigr\}{\partial^2\over \partial k^{\mu}\partial k^{\nu}}S_{ji}(k)\Bigr|_{k=xp}.
\label{C_1_2}
\eeq
We consider the last and most complicated diagram, the right-hand diagram in 
Fig. \ref{quark_diagrammatic} which is given by
\beq
W^{\rm Fig.1right}&=&{1\over 2}\int{d^4k_1\over (2\pi)^4}\int{d^4k_2\over (2\pi)^4}
\int{d^4k_3\over (2\pi)^4}
\int d^4\xi\,e^{ik_1\cdot \xi}\int d^4\eta\,e^{i(k_3-k_1)\cdot \eta}
\int d^4\chi\,e^{i(k_2-k_3)\cdot \chi}
\nonumber\\
&&\times\la P|\bar{\psi}_{jj'}(0)gA_b^{\nu}(\chi)gA_a^{\mu}(\eta)\psi_{ii'}(\xi)|P\ra 
S^{ab}_{\mu\nu\,ji:j'i'}(k_1,k_2,k_3).
\eeq
The factor $1/2$ is needed due to the interchange symmetry of the two gluon lines.
There are six diagrams in $S^{ab}_{\mu\nu\,ji:j'i'}(k_1,k_2,k_3)$.
\begin{figure}[H]
\begin{center}
  \includegraphics[height=8cm,width=16cm]{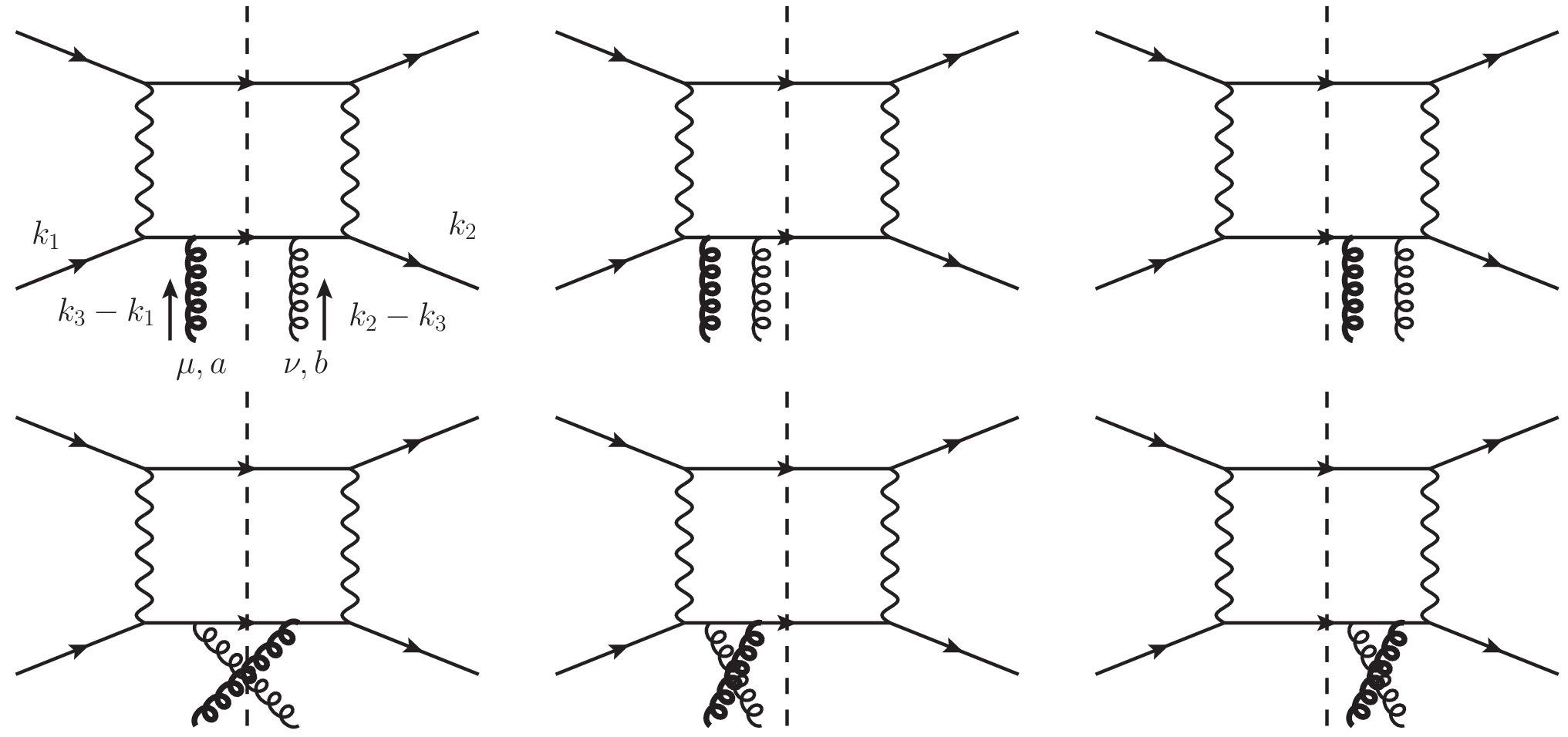}
\end{center}
 \caption{Hard scattering diagrams in $S^{ab}_{\mu\nu\,ji:j'i'}(k_1,k_2,k_3)$. The thick gluon line 
 represents the gluon line with the momentum $k_3-k_1$ and indices $\mu$ and $a$.}
\end{figure}
\noindent
These diagrams satisfy the WTIs
\beq
(k_3-k_1)^{\mu}S_{\mu\nu\,ji:j'i'}^{ab}(k_1,k_2,k_3)
&=&(T^bT^a)_{j'i'}\Bigl(
S_{\nu\,ji}(k_3,k_2)-S_{\nu\,ji}(k_1,k_2)\Bigr)
\nonumber\\
&&+(T^aT^b)_{j'i'}\Bigl(
-S_{\nu\,ji}(k_1,k_1+k_2-k_3)+S_{\nu\,ji}(k_1,k_2)\Bigr),
\label{WTI2}\\
(k_2-k_3)^{\nu}S_{\mu\nu\,ji:j'i'}^{ab}(k_1,k_2,k_3)&=&(T^bT^a)_{j'i'}\Bigl(
-S_{\mu\,ji}(k_1,k_3)+S_{\mu\,ji}(k_1,k_2)\Bigr)
\nonumber\\
&&+(T^aT^b)_{j'i'}\Bigl(
S_{\mu\,ji}(k_1+k_2-k_3,k_2)-S_{\mu\,ji}(k_1,k_2)\Bigr).
\label{WTI3}
\eeq
We first use the WTI (\ref{WTI2}) which gives an identity corresponding to (\ref{WTI_identity1}) as
\beq
A_a^{\mu}(\eta)S_{\mu\nu\,ji:j'i'}^{ab}(k_1,k_2,k_3)
&=&
-{1\over (k_3-k_1)\cdot n-i\epsilon}\omega^{\mu}_{\ \alpha}\Bigl[
(k_3-k_1)^{\alpha}A_a^n(\eta)-(k_3-k_1)\cdot n A_a^{\alpha}(\eta)
\Bigr]S_{\mu\nu\,ji:j'i'}^{ab}(k_1,k_2,k_3)
\nonumber\\
&&+{1\over (k_3-k_1)\cdot n-i\epsilon}A_a^n(\eta)\Bigl[
(T^bT^a)_{j'i'}\Bigl(
S_{\nu\,ji}(k_3,k_2)-S_{\nu\,ji}(k_1,k_2)\Bigr)
\nonumber\\
&&+(T^aT^b)_{j'i'}\Bigl(
-S_{\nu\,ji}(k_1,k_1+k_2-k_3)+S_{\nu\,ji}(k_1,k_2)\Bigr)\Bigr].
\eeq
Using this identity, $W^{\rm Fig.1right}$ is rewritten as
\beq
W^{\rm Fig.1right}&=&{1\over 2}\int{d^4k_1\over (2\pi)^4}\int{d^4k_2\over (2\pi)^4}
\int{d^4k_3\over (2\pi)^4}
\int d^4\xi\,e^{ik_1\cdot \xi}\int d^4\eta\,e^{i(k_3-k_1)\cdot \eta}
\int d^4\chi\,e^{i(k_2-k_3)\cdot \chi}
\nonumber\\
&&\times{1\over (k_3-k_1)\cdot n-i\epsilon}\Bigl[-i\omega^{\mu}_{\ \alpha}
\la P|\bar{\psi}_{jj'}(0)gA_b^{\nu}(\chi)gF_a^{(0)\alpha n}(\eta)\psi_{ii'}(\xi)|P\ra 
S^{ab}_{\mu\nu\,ji:j'i'}(k_1,k_2,k_3)
\nonumber\\
&&+\la P|\bar{\psi}_{j}(0)gA^{\nu}(\chi)gA^{n}(\eta)\psi_{i}(\xi)|P\ra 
\Bigl(S_{\nu\,ji}(k_3,k_2)-S_{\nu\,ji}(k_1,k_2)\Bigr)
\nonumber\\
&&+\la P|\bar{\psi}_{j}(0)gA^{n}(\eta)gA^{\nu}(\chi)\psi_{i}(\xi)|P\ra 
\Bigl(-S_{\nu\,ji}(k_1,k_1+k_2-k_3)+S_{\nu\,ji}(k_1,k_2)\Bigr)
\Bigr].
\eeq
We next use the other WTI (\ref{WTI3}) which gives
\beq
p^{\nu}S_{\mu\nu\,ji:j'i'}^{ab}(k_1,k_2,k_3)&=&
{1\over (k_2-k_3)\cdot n-i\epsilon}\Bigl[
-\omega^{\nu}_{\ \beta}(k_2-k_3)^{\beta}S_{\mu\nu\,ji:j'i'}^{ab}(k_1,k_2,k_3)
\nonumber\\
&&+(T^bT^a)_{j'i'}\Bigl(
-S_{\mu\,ji}(k_1,k_3)+S_{\mu\,ji}(k_1,k_2)\Bigr)
\nonumber\\
&&+(T^aT^b)_{j'i'}\Bigl(
S_{\mu\,ji}(k_1+k_2-k_3,k_2)-S_{\mu\,ji}(k_1,k_2)\Bigr)\Bigr],
\eeq
We need the WTI for $S_{\mu\,ji}(k_1,k_2)$ given in (\ref{WTI1}),
\beq
p^{\nu}S_{\nu\,ji}(k_3,k_2)&=&{1\over (k_2-k_3)\cdot n-i\epsilon}\Bigl[
-\omega^{\nu}_{\ \beta}(k_2-k_3)^{\beta}S_{\nu\,ji}(k_3,k_2)
+S_{ji}(k_2)-S_{ji}(k_3)\Bigr],
\\
p^{\nu}S_{\nu\,ji}(k_1,k_2)&=&{1\over (k_2-k_1)\cdot n-i\epsilon}\Bigl[
-\omega^{\nu}_{\ \beta}[(k_2-k_3)^{\beta}+(k_3-k_1)^{\beta}]S_{\nu\,ji}(k_1,k_2)
+S_{ji}(k_2)-S_{ji}(k_1)\Bigr],
\eeq
\beq
&&p^{\nu}S_{\nu\,ji}(k_1,k_1+k_2-k_3)
\nonumber\\
&=&{1\over (k_2-k_3)\cdot n-i\epsilon}\Bigl[
-\omega^{\nu}_{\ \beta}(k_2-k_3)^{\beta}S_{\nu\,ji}(k_1,k_1+k_2-k_3)
+S_{ji}(k_1+k_2-k_3)-S_{ji}(k_1)\Bigr],
\eeq
Using these identities, we obtain
\beq
&&W^{\rm Fig.1right}
\nonumber\\
&=&\int{d^4k_1\over (2\pi)^4}\int{d^4k_2\over (2\pi)^4}
\int{d^4k_3\over (2\pi)^4}
\int d^4\xi\,e^{ik_1\cdot \xi}\int d^4\eta\,e^{i(k_3-k_1)\cdot \eta}
\int d^4\chi\,e^{i(k_2-k_3)\cdot \chi}
\nonumber\\
&&\times\Bigl\{
-{1\over 2}\omega^{\mu}_{\ \alpha}\omega^{\nu}_{\ \beta}
\la P|\bar{\psi}_{jj'}(0)gF_b^{(0)\beta n}(\chi)gF_a^{(0)\alpha n}(\eta)\psi_{ii'}(\xi)|P\ra 
{1\over (k_3-k_1)\cdot n-i\epsilon}{1\over (k_2-k_3)\cdot n-i\epsilon}S^{ab}_{\mu\nu\,ji:j'i'}(k_1,k_2,k_3)
\nonumber\\
&&+{1\over (k_3-k_1)\cdot n-i\epsilon}{1\over (k_2-k_3)\cdot n-i\epsilon}
\Bigl[i\omega^{\mu}_{\ \alpha}
\la P|\bar{\psi}_{j}(0)gA^n(\chi)gF^{(0)\alpha n}(\eta)\psi_{i}(\xi)|P\ra S_{\mu\,ji}(k_1,k_3)
\nonumber\\
&&-i\omega^{\mu}_{\ \alpha}\la P|\bar{\psi}_{j}(0)gF^{(0)\alpha n}(\chi)gA^{n}(\eta)\psi_{i}(\xi)|P\ra 
S_{\mu\,ji}(k_3,k_2)\Bigr]
\nonumber\\
&&+\Bigl[-{i\over (k_2-k_3)\cdot n-i\epsilon}
\omega^{\mu}_{\ \alpha}
\la P|\bar{\psi}_{j}(0)gA^n(\chi)gF^{(0)\alpha n}(\eta)\psi_{i}(\xi)|P\ra
\nonumber\\
&&+{i\over (k_3-k_1)\cdot n-i\epsilon}
\omega^{\mu}_{\ \alpha}\la P|\bar{\psi}_{j}(0)gF^{(0)\alpha n}(\chi)gA^{n}(\eta)\psi_{i}(\xi)|P\ra
\nonumber\\
&&+\omega^{\mu}_{\ \alpha}\la P|\bar{\psi}_{j}(0)gA^{n}(\chi)gA^{\alpha}(\eta)\psi_{i}(\xi)|P\ra 
-\omega^{\mu}_{\ \alpha}\la P|\bar{\psi}_{j}(0)gA^{\alpha}(\chi)gA^{n}(\eta)\psi_{i}(\xi)|P\ra 
\Bigr]{1\over (k_2-k_1)\cdot n-i\epsilon}S_{\mu\,ji}(k_1,k_2)
\nonumber\\
&&+\la P|\bar{\psi}_{j}(0)gA^n(\chi)gA^{n}(\eta)\psi_{i}(\xi)|P\ra 
\Bigl({1\over (k_3-k_1)\cdot n-i\epsilon}{1\over (k_2-k_1)\cdot n-i\epsilon}S_{ji}(k_1)
\nonumber\\
&&+{1\over (k_2-k_3)\cdot n-i\epsilon}{1\over (k_2-k_1)\cdot n-i\epsilon} S_{ji}(k_2)
-{1\over (k_3-k_1)\cdot n-i\epsilon}{1\over (k_2-k_3)\cdot n-i\epsilon} S_{ji}(k_3)\Bigr)
\Bigr\},
\eeq
where we appropriately changed the integration variables to $k_3\to k_1+k_2-k_3$ and
$\eta\leftrightarrow\chi$.
We pick up the first term in the Taylor expansion of $S^{ab}_{\mu\nu\,ji:j'i'}(k_1,k_2,k_3)$
for the first term in $W^{\rm Fig.1right}$,
\beq
&&-{1\over 2}\omega^{\mu}_{\ \alpha}\omega^{\nu}_{\ \beta}
\int dx_1\int dx_2\int dx_3\int{d\lambda\over 2\pi}\int{d\mu\over 2\pi}\int{d\omega\over 2\pi}
e^{i\lambda x_1}e^{i\mu (x_3-x_1)}e^{i\omega (x_2-x_3)}
\nonumber\\
&&\times
\la P|\bar{\psi}_{jj'}(0)gF_b^{(0)\beta n}(\omega n)gF_a^{(0)\alpha n}(\mu n)\psi_{ii'}(\lambda n)|P\ra 
{1\over x_3-x_1-i\epsilon}{1\over x_2-x_3-i\epsilon}S^{ab}_{\mu\nu\,ji:j'i'}(x_1,x_2,x_3).
\label{C_4}
\eeq
We pick up the first derivative terms in the Taylor expansion for $S_{\mu\,ji}(k_i,k_j)$,
and then obtain
\beq
&&\omega^{\mu}_{\ \alpha}\omega^{\nu}_{\ \beta}\int{dx_1}\int{dx_2}
\int {d\lambda\over 2\pi}\,e^{i\lambda x_1}\int {d\mu\over 2\pi}\eta\,e^{i\mu(x_2-x_1)}
\nonumber\\
&&\Bigl\{\Bigl[
\la P|\bar{\psi}_{j}(0)\Bigl(ig\int^{0}_{\mu} dt\,A^n(t n)\Bigr)
gF^{(0)\alpha n}(\mu n)\partial^{\beta}\psi_{i}(\lambda n)|P\ra
+i\int^{\infty}_{\lambda}dt
\la P|\bar{\psi}_{j}(0)gF^{(0)\alpha n}(\mu n)gF^{(0)\beta n}(t n)\psi_{i}(\lambda n)|P\ra 
\nonumber\\
&&+\la P|\bar{\psi}_{j}(0)gF^{(0)\alpha n}(\mu n)\Bigl(-igA^{\beta}(\lambda n)\Bigr)
\psi_{i}(\lambda n)|P\ra 
+\la P|\bar{\psi}_{j}(0)gF^{(0)\alpha n}(\mu n)
\Bigl(ig\int^{\mu}_{\lambda}dt\,A^{n}(t n)\Bigr)\partial^{\beta}\psi_{i}(\lambda n)|P\ra 
\nonumber\\
&&+\la P|\bar{\psi}_{j}(0)\Bigl(-ig^2[A^{\alpha}(\mu n),A^{n}(\mu n)]\Bigr)
\partial^\beta\psi_{i}(\lambda n)|P\ra 
\Bigr]{1\over x_2-x_1-i\epsilon}{\partial\over \partial k_1^{\nu}}S_{\mu\,ji}(k_1,k_2)\Bigr|_{k_i=x_ip}
\nonumber\\
&&\Bigl[\la P|\bar{\psi}_{j}(0)\Bigl(ig\int^{0}_{\mu}dt\,A^n(t n)\Bigr)
\partial^{\beta}\Bigl(gF^{(0)\alpha n}(\mu n)\psi_{i}(\lambda n)\Bigr)|P\ra
+\la P|\bar{\psi}_{j}(0)\partial^{\beta}\Bigl[gF^{(0)\alpha n}(\mu n)
\Bigl(ig\int^{\mu}_{\lambda}dt\,A^{n}(t n)\Bigr)\psi_{i}(\lambda n)\Bigr]|P\ra 
\nonumber\\
&&+i\int^{\infty}_{\mu}dt\,\la P|\bar{\psi}_{j}(0)gF^{(0)\beta n}(t n)
gF^{(0)\alpha n}(\mu n)\psi_{i}(\lambda n)|P\ra
+\la P|\bar{\psi}_{j}(0)\Bigl(-igA^{\beta}(\mu n)\Bigr)gF^{(0)\alpha n}(\mu n)\psi_{i}(\lambda n)|P\ra
\nonumber\\
&&+\la P|\bar{\psi}_{j}(0)\partial^{\beta}\Bigl(-ig^2[A^{\alpha}(\mu n),A^{n}(\mu n)]
\psi_{i}(\lambda n)\Bigr)|P\ra
\Bigr]{1\over x_2-x_1-i\epsilon}{\partial\over \partial k_2^{\nu}}S_{\mu\,ji}(k_1,k_2)\Bigr|_{k_i=x_ip}
\Bigr\}.
\label{C_2}
\eeq
We pick up the second derivative terms in the Taylor expansion for $S_{ji}(k_i)$,
and then obtain
\beq
&&-{1\over 2}\omega^{\mu}_{\ \alpha}\omega^{\nu}_{\ \beta}\int{dx}
\int {d\lambda\over 2\pi}\,e^{i\lambda x}
\Bigl\{\la P|\bar{\psi}_{j}(0)\Bigl((ig)^2\int^{0}_{\lambda} dt\int^{t}_{\lambda} dt'\,
A^n(t n)A^{n}(t' n)\Bigr)
\partial^{\beta}\partial^{\alpha}
\psi_{i}(\lambda n)|P\ra
\nonumber\\
&&+\la P|\bar{\psi}_{j}(0)\Bigl(ig\int^{0}_{\lambda}d\omega\,A^{n}(\omega n)\Bigr)
\Bigl(-igA^{\beta}(\lambda n)\Bigr)
\partial^{\alpha}\psi_{i}(\lambda n)|P\ra
\nonumber\\
&&+\la P|\bar{\psi}_{j}(0)\Bigl(ig\int^{0}_{\lambda}d\omega\,A^{n}(\omega n)\Bigr)
\partial^{\beta}\Bigl[
\Bigl(-igA^{\alpha}(\lambda n)\Bigr)\psi_{i}(\lambda n)\Bigr]|P\ra
\nonumber\\
&&+\la P|\bar{\psi}_{j}(0)\Bigl(-igA^{\beta}(\lambda n)\Bigr)
\Bigl(-igA^{\alpha}(\lambda n)\Bigr)\psi_{i}(\lambda n)|P\ra
\nonumber\\
&&+i\int^{\infty}_{\lambda} d\mu
\la P|\bar{\psi}_{j}(0)\Bigl(ig\int^{0}_{\mu} d\omega A^{n}(\omega n)\Bigr)
\partial^{\beta}\Bigl(gF^{(0)\alpha n}(\mu n)\psi_{i}(\lambda n)\Bigr)|P\ra
\nonumber\\
&&+i\int^{\infty}_{\lambda}d\mu
\la P|\bar{\psi}_{j}(0)\Bigl(ig\int^{0}_{\mu}d\omega\,A^{n}(\omega n)\Bigr)gF^{(0)\beta n}(\mu n)
\partial^{\alpha}\psi_{i}(\lambda n)|P\ra
\nonumber\\
&&+i\int^{\infty}_{\lambda}d\omega
\la P|\bar{\psi}_{j}(0)\partial^{\beta}\Bigl[
gF^{(0)\alpha n}(\omega n)\Bigl(ig\int^{\omega}_{\lambda}d\mu\,A^n(\mu n)\Bigr)
\psi_{i}(\lambda n)\Bigr]|P\ra
\nonumber\\
&&-ig\int^{\infty}_{\lambda}d\mu\,\la P|\bar{\psi}_{j}(0)\partial^{\beta}\Bigl[
\Bigl(igA^{\alpha}(\mu n)A^n(\mu n)\Bigr)\psi_{i}(\lambda n)\Bigr]|P\ra
\nonumber\\
&&-\int^{\infty}_{\lambda} d\mu\int^{\infty}_{\mu} d\omega
\la P|\bar{\psi}_{j}(0)gF^{(0)\beta n}(\omega n)gF^{(0)\alpha n}(\mu n)\psi_{i}(\lambda n)|P\ra
\nonumber\\
&&+i\int^{\infty}_{\lambda}d\mu\,
\la P|\bar{\psi}_{j}(0)\Bigl(-igA^{\beta}(\mu n)\Bigr)
gF^{(0)\alpha n}(\mu n)\psi_{i}(\lambda n)|P\ra
\nonumber\\
&&+i\int^{\infty}_{\lambda}d\omega\,
\la P|\bar{\psi}_{j}(0)gF^{(0)\beta n}(\omega n)
\Bigl(-igA^{\alpha}(\lambda n)\Bigr)\psi_{i}(\lambda n)|P\ra
\nonumber\\
&&+ig\int^{\infty}_{\lambda}d\mu\,\la P|\bar{\psi}_{j}(0)\partial^{\beta}\Bigl(
igA^{n}(\mu n)A^{\alpha}(\mu n)\psi_{i}(\lambda n)\Bigr)|P\ra
\nonumber\\
&&+ig\int^{\infty}_{\lambda}d\mu\,
\la P|\bar{\psi}_{j}(0)\Bigl(-igA^{\beta}(\mu n)A^n(\mu n)\Bigr)
\partial^{\alpha}\psi_{i}(\lambda n)|P\ra
\nonumber\\
&&+i\int^{\infty}_{\lambda}d\omega
\la P|\bar{\psi}_{j}(0)gF^{(0)\beta n}(\omega n)\Bigr(ig\int^{\omega}_{\lambda}d\mu A^n(\mu n)\Bigr)
\partial^{\alpha}\psi_{i}(\lambda n)|P\ra
\nonumber\\
&&+ig\int^{\infty}_{\lambda}d\mu\,
\la P|\bar{\psi}_{j}(0)\Bigl(igA^{n}(\mu n)A^{\beta}(\mu n)\Bigr)
\partial^{\alpha}\psi_{i}(\lambda n)|P\ra
\Bigr\}{\partial^2\over \partial k^{\mu}\partial k^{\nu}}S_{ji}(k)\Bigr|_{k=xp}
\label{C_1_3}
\eeq
Our calculation shows that the twist-4 contributions extracted from the diagrams in 
Fig. \ref{quark_diagrammatic} take the following form.
\beq
&&-{1\over 2}\omega^{\mu}_{\ \alpha}\omega^{\nu}_{\ \beta}\int dx\,C^{\alpha\beta}_{1ij}(x)
{\partial^2\over \partial k^{\mu}\partial k^{\nu}}S_{ji}(k)\Bigr|_{k=xp}
\nonumber\\
&&+\omega^{\mu}_{\ \alpha}\omega^{\nu}_{\ \beta}\int {dx_1}\int {dx_2}\,
{1\over x_2-x_1-i\epsilon}
\,\Bigl(C^{\alpha\beta}_{2ij}(x_1,x_2)
{\partial\over \partial k_1^{\nu}}S_{\mu\,ji}(k_1,k_2)\Bigr|_{k_i=x_ip}
+C^{\alpha\beta}_{3ij}(x_1,x_2)
{\partial\over \partial k_2^{\nu}}S_{\mu\,ji}(k_1,k_2)\Bigr|_{k_i=x_ip}\Bigr)
\nonumber\\
&&-{1\over 2}\omega^{\mu}_{\ \alpha}\omega^{\nu}_{\ \beta}\int {dx_1}\int {dx_2}\int {dx_3}\,
{1\over x_3-x_1-i\epsilon}{1\over x_2-x_3-i\epsilon}C^{\alpha\beta}_{4\,ab\,ij:i'j'}(x_1,x_2,x_3)
S^{ab}_{\mu\nu\,ji:j'i'}(x_1,x_2,x_3).
\label{twist-4_form_quark}
\eeq
We find that $C_{2,3,4}$ are the matrix elements appearing in the gluon condensate 
term (\ref{gluon_condensate_expansion}).
$C^{\alpha\beta}_{4\,ab\,ij:i'j'}$ is given by (\ref{C_4}),
\beq
&&C^{\alpha\beta}_{4\,ab\,ij:i'j'}(x_1,x_2,x_3)
\nonumber\\
&=&\int{d\lambda\over 2\pi}\int{d\mu\over 2\pi}\int{d\omega\over 2\pi}
e^{i\lambda x_1}e^{i\mu (x_3-x_1)}e^{i\omega (x_2-x_3)}
\la P|\bar{\psi}_{jj'}(0)gF_b^{(0)\beta n}(\omega n)gF_a^{(0)\alpha n}(\mu n)\psi_{ii'}(\lambda n)|P\ra
\nonumber\\
&=&M^{Q\,\alpha\beta}_{FF\,ab\,ij:i'j'}(x_1,x_2,x_3)\Bigr|_{{\rm up\ to\ }g^2},
\eeq
which gives (\ref{quark_FF}) when it is contracted with $(T^bT^a)_{i'j'}$ in 
$S^{ab}_{\mu\nu\,ji:j'i'}(x_1,x_2,x_3)$.
$C^{\alpha\beta}_{2\,ij}$ and $C^{\alpha\beta}_{3\,ij}$  are
given by (\ref{C_2_1}) and (\ref{C_2}),
\beq
&&C^{\alpha\beta}_{2ij}(x_1,x_2)
\nonumber\\
&=&\int{d\lambda\over 2\pi}\int{d\mu\over 2\pi}
e^{i\lambda x_1}e^{i\mu (x_2-x_1)}
\Bigl[
\la P|\bar{\psi}_{j}(0)gF^{(0)\alpha n}(\mu n)\partial^{\beta}\psi_{i}(\lambda n)|P\ra 
\nonumber\\
&&+\la P|\bar{\psi}_{j}(0)\Bigl(ig\int^{0}_{\mu} dt\,A^n(t n)\Bigr)
gF^{(0)\alpha n}(\mu n)\partial^{\beta}\psi_{i}(\lambda n)|P\ra
\nonumber\\
&&+i\int^{\infty}_{\lambda}dt
\la P|\bar{\psi}_{j}(0)gF^{(0)\alpha n}(\mu n)gF^{(0)\beta n}(t n)\psi_{i}(\lambda n)|P\ra 
\nonumber\\
&&+\la P|\bar{\psi}_{j}(0)gF^{(0)\alpha n}(\mu n)\Bigl(-igA^{\beta}(\lambda n)\Bigr)
\psi_{i}(\lambda n)|P\ra 
+\la P|\bar{\psi}_{j}(0)gF^{(0)\alpha n}(\mu n)
\Bigl(ig\int^{\mu}_{\lambda}dt\,A^{n}(t n)\Bigr)\partial^{\beta}\psi_{i}(\lambda n)|P\ra 
\nonumber\\
&&+\la P|\bar{\psi}_{j}(0)\Bigl(-ig^2[A^{\alpha}(\mu n),A^{n}(\mu n)]\Bigr)
\partial^\beta\psi_{i}(\lambda n)|P\ra\Bigr]
=M^{Q\,\alpha\beta}_{1\partial ij}(x_1,x_2)\Bigr|_{{\rm up\ to\ }g^2}.
\eeq
\beq
&&C^{\alpha\beta}_{3ij}(x_1,x_2)
\nonumber\\
&=&\int{d\lambda\over 2\pi}\int{d\mu\over 2\pi}
e^{i\lambda x_1}e^{i\mu (x_2-x_1)}
\Bigl[
\la P|\bar{\psi}_{j}(0)g\partial^{\beta}\Bigl(F^{(0)\alpha n}(\mu n)\psi_{i}(\lambda n)\Bigr)|P\ra
\nonumber\\
&&+\la P|\bar{\psi}_{j}(0)\Bigl(ig\int^{0}_{\mu}dt\,A^n(t n)\Bigr)
\partial^{\beta}\Bigl(gF^{(0)\alpha n}(\mu n)\psi_{i}(\lambda n)\Bigr)|P\ra
\nonumber\\
&&+\la P|\bar{\psi}_{j}(0)\partial^{\beta}\Bigl[gF^{(0)\alpha n}(\mu n)
\Bigl(ig\int^{\mu}_{\lambda}dt\,A^{n}(t n)\Bigr)\psi_{i}(\lambda n)\Bigr]|P\ra 
\nonumber\\
&&+i\int^{\infty}_{\mu}dt\,\la P|\bar{\psi}_{j}(0)gF^{(0)\beta n}(t n)
gF^{(0)\alpha n}(\mu n)\psi_{i}(\lambda n)|P\ra
+\la P|\bar{\psi}_{j}(0)\Bigl(-igA^{\beta}(\mu n)\Bigr)gF^{(0)\alpha n}(\mu n)\psi_{i}(\lambda n)|P\ra
\nonumber\\
&&+\la P|\bar{\psi}_{j}(0)\partial^{\beta}\Bigl(-ig^2[A^{\alpha}(\mu n),A^{n}(\mu n)]
\psi_{i}(\lambda n)\Bigr)|P\ra
\Bigr]=M^{Q\,\alpha\beta}_{2\partial ij}(x_1,x_2)\Bigr|_{{\rm up\ to\ }g^2}.
\eeq
$C^{\alpha\beta}_{1ij}$ does not appear in the gluon condensate term.
It is  constructed from (\ref{C_1_1}), (\ref{C_1_2}) and (\ref{C_1_3}),
\beq
&&C^{\alpha\beta}_{1ij}(x)
\nonumber\\
&=&\int{d\lambda\over 2\pi}e^{i\lambda x}\Bigl[\la P|\bar{\psi}_j(0)
\partial^{\beta}\partial^{\alpha}\psi_i(\lambda n)|P\ra
+i\int^{\infty}_{\lambda}d\mu\,\la P|\bar{\psi}_{j}(0)g
\partial^{\beta}\Bigl(F^{(0)\alpha n}(\mu n)\psi_{i}(\lambda n)\Bigr)|P\ra 
\nonumber\\
&&+\la P|\bar{\psi}_{j}(0)\partial^{\beta}\Bigl(-igA^{\alpha}(\lambda n)\psi_{i}(\lambda n)\Bigr)|P\ra 
+i\int^{\infty}_{\lambda}d\mu\,\la P|\bar{\psi}_{j}(0)g
F^{(0)\beta n}(\mu n)\partial^{\alpha}\psi_{i}(\lambda n)|P\ra 
\nonumber\\
&&+\la P|\bar{\psi}_{j}(0)\Bigl(-igA^{\beta}(\lambda n)\Bigr)\partial^{\alpha}\psi_{i}(\lambda n)|P\ra 
+\la P|\bar{\psi}_{j}(0)\Bigl(ig\int^0_{\lambda}d\mu\,A^n(\mu n)\Bigr)
\partial^{\beta}\partial^{\alpha}\psi_{i}(\lambda n)|P\ra 
\nonumber\\
&&+\la P|\bar{\psi}_{j}(0)\Bigl((ig)^2\int^{0}_{\lambda} dt\int^{t}_{\lambda} dt'\,
A^n(t n)A^{n}(t' n)\Bigr)
\partial^{\beta}\partial^{\alpha}
\psi_{i}(\lambda n)|P\ra
\nonumber\\
&&+\la P|\bar{\psi}_{j}(0)\Bigl(ig\int^{0}_{\lambda}d\omega\,A^{n}(\omega n)\Bigr)
\Bigl(-igA^{\beta}(\lambda n)\Bigr)
\partial^{\alpha}\psi_{i}(\lambda n)|P\ra
\nonumber\\
&&+\la P|\bar{\psi}_{j}(0)\Bigl(ig\int^{0}_{\lambda}d\omega\,A^{n}(\omega n)\Bigr)
\partial^{\beta}\Bigl[
\Bigl(-igA^{\alpha}(\lambda n)\Bigr)\psi_{i}(\lambda n)\Bigr]|P\ra
\nonumber\\
&&+\la P|\bar{\psi}_{j}(0)\Bigl(-igA^{\beta}(\lambda n)\Bigr)
\Bigl(-igA^{\alpha}(\lambda n)\Bigr)\psi_{i}(\lambda n)|P\ra
\nonumber\\
&&+i\int^{\infty}_{\lambda} d\mu
\la P|\bar{\psi}_{j}(0)\Bigl(ig\int^{0}_{\mu} d\omega A^{n}(\omega n)\Bigr)
\partial^{\beta}\Bigl(gF^{(0)\alpha n}(\mu n)\psi_{i}(\lambda n)\Bigr)|P\ra
\nonumber\\
&&+i\int^{\infty}_{\lambda}d\mu
\la P|\bar{\psi}_{j}(0)\Bigl(ig\int^{0}_{\mu}d\omega\,A^{n}(\omega n)\Bigr)gF^{(0)\beta n}(\mu n)
\partial^{\alpha}\psi_{i}(\lambda n)|P\ra
\nonumber\\
&&+i\int^{\infty}_{\lambda}d\omega
\la P|\bar{\psi}_{j}(0)\partial^{\beta}\Bigl[
gF^{(0)\alpha n}(\omega n)\Bigl(ig\int^{\omega}_{\lambda}d\mu\,A^n(\mu n)\Bigr)
\psi_{i}(\lambda n)\Bigr]|P\ra
\nonumber\\
&&-ig\int^{\infty}_{\lambda}d\mu\,\la P|\bar{\psi}_{j}(0)\partial^{\beta}\Bigl[
\Bigl(igA^{\alpha}(\mu n)A^n(\mu n)\Bigr)\psi_{i}(\lambda n)\Bigr]|P\ra
\nonumber\\
&&-\int^{\infty}_{\lambda} d\mu\int^{\infty}_{\mu} d\omega
\la P|\bar{\psi}_{j}(0)gF^{(0)\beta n}(\omega n)gF^{(0)\alpha n}(\mu n)\psi_{i}(\lambda n)|P\ra
\nonumber\\
&&+i\int^{\infty}_{\lambda}d\mu\,
\la P|\bar{\psi}_{j}(0)\Bigl(-igA^{\beta}(\mu n)\Bigr)
gF^{(0)\alpha n}(\mu n)\psi_{i}(\lambda n)|P\ra
\nonumber\\
&&+i\int^{\infty}_{\lambda}d\omega\,
\la P|\bar{\psi}_{j}(0)gF^{(0)\beta n}(\omega n)
\Bigl(-igA^{\alpha}(\lambda n)\Bigr)\psi_{i}(\lambda n)|P\ra
\nonumber\\
&&+ig\int^{\infty}_{\lambda}d\mu\,\la P|\bar{\psi}_{j}(0)\partial^{\beta}\Bigl(
igA^{n}(\mu n)A^{\alpha}(\mu n)\psi_{i}(\lambda n)\Bigr)|P\ra
\nonumber\\
&&+ig\int^{\infty}_{\lambda}d\mu\,
\la P|\bar{\psi}_{j}(0)\Bigl(-igA^{\beta}(\mu n)A^n(\mu n)\Bigr)
\partial^{\alpha}\psi_{i}(\lambda n)|P\ra
\nonumber\\
&&+i\int^{\infty}_{\lambda}d\omega
\la P|\bar{\psi}_{j}(0)gF^{(0)\beta n}(\omega n)\Bigr(ig\int^{\omega}_{\lambda}d\mu A^n(\mu n)\Bigr)
\partial^{\alpha}\psi_{i}(\lambda n)|P\ra
\nonumber\\
&&+ig\int^{\infty}_{\lambda}d\mu\,
\la P|\bar{\psi}_{j}(0)\Bigl(igA^{n}(\mu n)A^{\beta}(\mu n)\Bigr)
\partial^{\alpha}\psi_{i}(\lambda n)|P\ra
\Bigr]=M^{Q\,\alpha\beta}_{\partial\partial ij}(x)\Bigr|_{{\rm up\ to\ }g^2},
\eeq
where $M^{Q\,\alpha\beta}_{\partial\partial}(x)$ is given by
\beq
M^{Q\,\alpha\beta}_{\partial\partial}(x)&=&\int{d\lambda\over 2\pi}\,e^{i\lambda x}
\la P|\bar{\psi}(0)[0,\lambda n]D_{\perp}^{\beta}(\lambda n)
D_{\perp}^{\alpha}(\lambda n)\psi(\lambda n)|P\ra
\nonumber\\
&&+i\int{d\lambda\over 2\pi}\,e^{i\lambda x}
\int^{\infty}_{\lambda}dt\,
\la P|\bar{\psi}(0)[0,t n]gF^{\beta n}(t n)[t n,\lambda n]
D_{\perp}^{\alpha}(\lambda n)\psi(\lambda n)|P\ra
\nonumber\\
&&+i\int{d\lambda\over 2\pi}\,e^{i\lambda x}
\int^{\infty}_{\lambda} dt\,
\la P|\bar{\psi}(0)[0,t n]D_{\perp}^{\beta}(t n)\Bigl(
gF^{\alpha n}(t n)[t n,\lambda n]\psi(\lambda n)\Bigr)|P\ra
\nonumber\\
&&-\int{d\lambda\over 2\pi}\,e^{i\lambda x}
\int^{\infty}_{\lambda} dt\int^{\infty}_{t}dt'\,
\la P|\bar{\psi}(0)[0,t' n]gF^{\beta n}(t' n)[t' n,t n]
gF^{\alpha n}(t n)[t n,\lambda n]\psi(\lambda n)|P\ra,
\eeq
which is expressed as a $k_T$-moment of quark TMD matrix element.
\beq
M^{Q\,\alpha\beta}_{\partial\partial}(x)
&=&-\int d^2k_T\,k_T^{\alpha}k_T^{\beta}M^{Q}(x,k_T),
\eeq
There is another type of twist-4 quark contribution involving a 4-quark matrix element
as shown in the following diagram.
\begin{figure}[H]
\begin{center}
  \includegraphics[height=4cm,width=6cm]{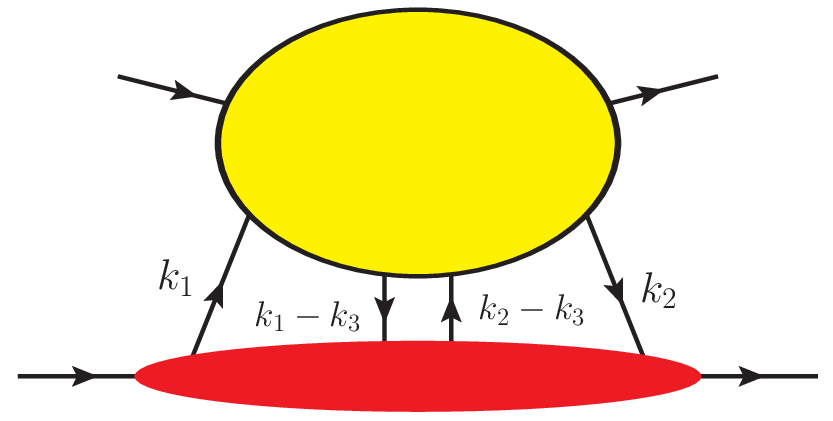}
\end{center}
 \caption{The diagram with four quark matrix element in the diagrammatic method.}
\label{4quark_diagrammatic}
\end{figure}
\noindent
This diagram is given by
\beq
W^{\rm Fig.5}&=&\int{d^4k_1\over (2\pi)^4}\int{d^4k_2\over (2\pi)^4}
\int{d^4k_3\over (2\pi)^4}
\int d^4\xi\,e^{ik_1\cdot \xi}\int d^4\eta\,e^{i(k_3-k_1)\cdot \eta}
\int d^4\chi\,e^{i(k_2-k_3)\cdot \chi}
\nonumber\\
&&\times\la P|\bar{\psi}_{jj'}(0)g\psi_{kk'}(\chi)g\bar{\psi}_{ll'}(\eta)\psi_{ii'}(\xi)|P\ra 
S_{ijkl:i'j'k'l'}(k_1,k_2,k_3).
\eeq
The twist-4 contribution is given by the first term in the Taylor expansion of $S_{ijkl:i'j'k'l'}(k_1,k_2,k_3)$
and we obtain
\beq
&&\int dx_1\int dx_2\int dx_3
\int{d\lambda\over 2\pi}\int{d\mu\over 2\pi}\int{d\omega\over 2\pi}\,
e^{i\lambda x_1}e^{i\mu(x_3-x_1)}e^{i\omega(x_2-x_3)}
\nonumber\\
&&\times\la P|\bar{\psi}_{jj'}(0)g\psi_{kk'}(\omega n)g\bar{\psi}_{ll'}(\mu n)\psi_{ii'}(\lambda n)|P\ra 
S_{ijkl:i'j'k'l'}(x_1p,x_2p,x_3p).
\eeq
$S_{ijkl:i'j'k'l'}(x_1p,x_2p,x_3p)$ could give two types of color structures
\beq
S_{ijkl:i'j'k'l'}(x_1p,x_2p,x_3p)=S^1_{ijkl}(x_1p,x_2p,x_3p)\delta^{i'l'}\delta^{k'j'}
+S^2_{ijkl}(x_1p,x_2p,x_3p)\delta^{i'j'}\delta^{k'l'}.
\eeq
We expect that the gauge-invariant matrix elements can be constructed as
\beq
\int dx_1\int dx_2\int dx_3\Bigl(M^{4Q}_{ik:jl}(x_1,x_2,x_3)S^1_{ijkl}(x_1p,x_2p,x_3p)
-M^{4Q}_{ki:jl}(x_2-x_3,x_2,x_2-x_1)S^2_{ijkl}(x_1p,x_2p,x_3p)\Bigr).
\eeq
We finally obtain the result of quark twist-4 contributions,
\beq
&&P_h^0{d\sigma\over d^3\vec{P}_h}
\nonumber\\
&=&{\alpha_{em}^2\over S}\int{dz\over z^2}D(z)
\Bigl\{
-{1\over 2}\omega^{\mu}_{\ \alpha}\omega^{\nu}_{\ \beta}\int dx\,
M^{Q\,\alpha\beta}_{\partial\partial\,ij}(x)
{\partial^2\over \partial k^{\mu}\partial k^{\nu}}S_{ji}(k)\Bigr|_{k=xp}
\nonumber\\
&&+\omega^{\mu}_{\ \alpha}\omega^{\nu}_{\ \beta}\int {dx_1}\int {dx_2}\,
{1\over x_2-x_1-i\epsilon}
\,\Bigl(M^{Q\,\alpha\beta}_{1\partial\,ij}(x_1,x_2)
{\partial\over \partial k_1^{\nu}}S_{\mu\,ji}(k_1,k_2)
+M^{Q\,\alpha\beta}_{2\partial\,ij}(x_1,x_2)
{\partial\over \partial k_2^{\nu}}S_{\mu\,ji}(k_1,k_2)\Bigr)\Bigr|_{k_i=x_ip}
\nonumber\\
&&-{1\over 2}\omega^{\mu}_{\ \alpha}\omega^{\nu}_{\ \beta}\int {dx_1}\int {dx_2}\int {dx_3}\,
{1\over x_3-x_1-i\epsilon}{1\over x_2-x_3-i\epsilon}M^{Q\,\alpha\beta}_{FF\,ab\,ij:i'j'}(x_1,x_2,x_3)
S^{ab}_{\mu\nu\,ji:j'i'}(x_1,x_2,x_3)
\nonumber\\
&&+\int dx_1\int dx_2\int dx_3\Bigl(M^{4Q}_{ik:jl}(x_1,x_2,x_3)S^1_{ijkl}(x_1p,x_2p,x_3p)
\nonumber\\
&&-M^{4Q}_{ki:jl}(x_2-x_3,x_2,x_2-x_1)S^2_{ijkl}(x_1p,x_2p,x_3p)\Bigr)
\Bigr\}.
\eeq
Note that we have to additionally take into account twist-4 components of 
the matrix elements composed of two or three operators, which have been extensively studied in
the context of transverse-spin asymmetries\cite{Kanazawa:2015ajw}, for the complete twist-4 result.
Since our derivation relies solely on WTIs which are generally satisfied in hard scattering,
the same cross section formula can be applied to many processes in $ep$ collisions.
Finding suitable processes to investigate the gluon condensate term is a desirable future task.


\section{Derivation of gluon twist-4 contributions to the unpolarized cross section in $ep$ collisions}

We here consider light hadron production in semi-inclusive deep inelastic scattering,
\beq
e(\ell)+p(P)\to e(\ell')+h(P_h)+X,
\eeq
where the unpolarized cross section receives gluon distribution contributions at leading-order
in the QCD coupling. The unpolarized cross section is written in terms of the following variables.
\beq
S_{ep}&=&(P+\ell)^2,\hspace{5mm}
Q^2=-q^2=-(\ell-\ell')^2,\hspace{5mm}
x_{bj}=\frac{Q^2}{2P\cdot q},\hspace{5mm}
z_f=\frac{P\cdot P_h}{P\cdot q}.
\eeq
The unpolarized cross section formula is given by
\beq
\frac{d^{6}\sigma}{dx_{bj}dQ^{2}dz_{f}dP^{2}_{h}d\phi d\chi}  =
\frac{\alpha ^{2}_{em}}{128\pi^4 z_fS^{2}_{ep} x^{2}_{bj} Q^{2}}
L^{\rho\sigma}(\ell,\ell')\int{dz\over z^2}D(z)W_{\rho\sigma}(P,q,{P_h\over z}), 
\eeq
where $L^{\rho\sigma}=2(\ell^{\rho}\ell^{\prime\sigma}
+\ell^{\sigma}\ell^{\prime\rho}-Q^2/2g^{\rho\sigma})$ is the leptonic tensor.
The hadronic tensor $W_{\rho\sigma}$ describes the interaction between
the proton and the virtual photon emitted by the electron.
We consider the following diagrams in order to derive the twist-4 contributions as in the case of the 
quark distribution.
\begin{figure}[H]
\begin{center}
  \includegraphics[height=4cm,width=16cm]{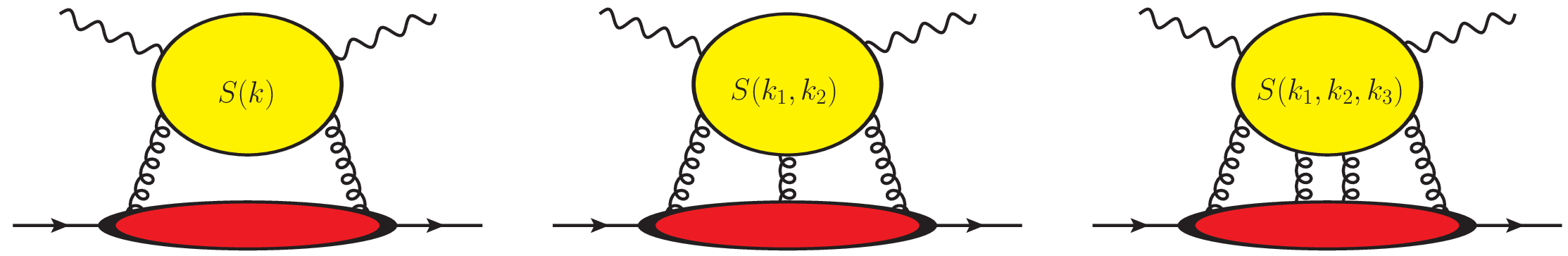}
\end{center}
 \caption{Gluon distribution contributions appear in the diagrammatic method.}
\label{gluon_diagrammatic}
\end{figure}
\noindent
We omit the Lorentz indices $\rho$ and $\sigma$ of the external virtual photon for simplicity below.
The left diagram is given by
\beq
W^{\rm Fig.6left}=\int{d^4k\over (2\pi)^4}\int d^4\xi\,e^{ik\cdot \xi}
\la P|A_b^{\nu}(0)A_a^{\mu}(\xi)|P\ra S^{ab}_{\mu\nu}(k).
\eeq
There are four diagrams in $S^{ab}_{\mu\nu}(k)$.
\begin{figure}[H]
\begin{center}
  \includegraphics[height=8cm,width=12cm]{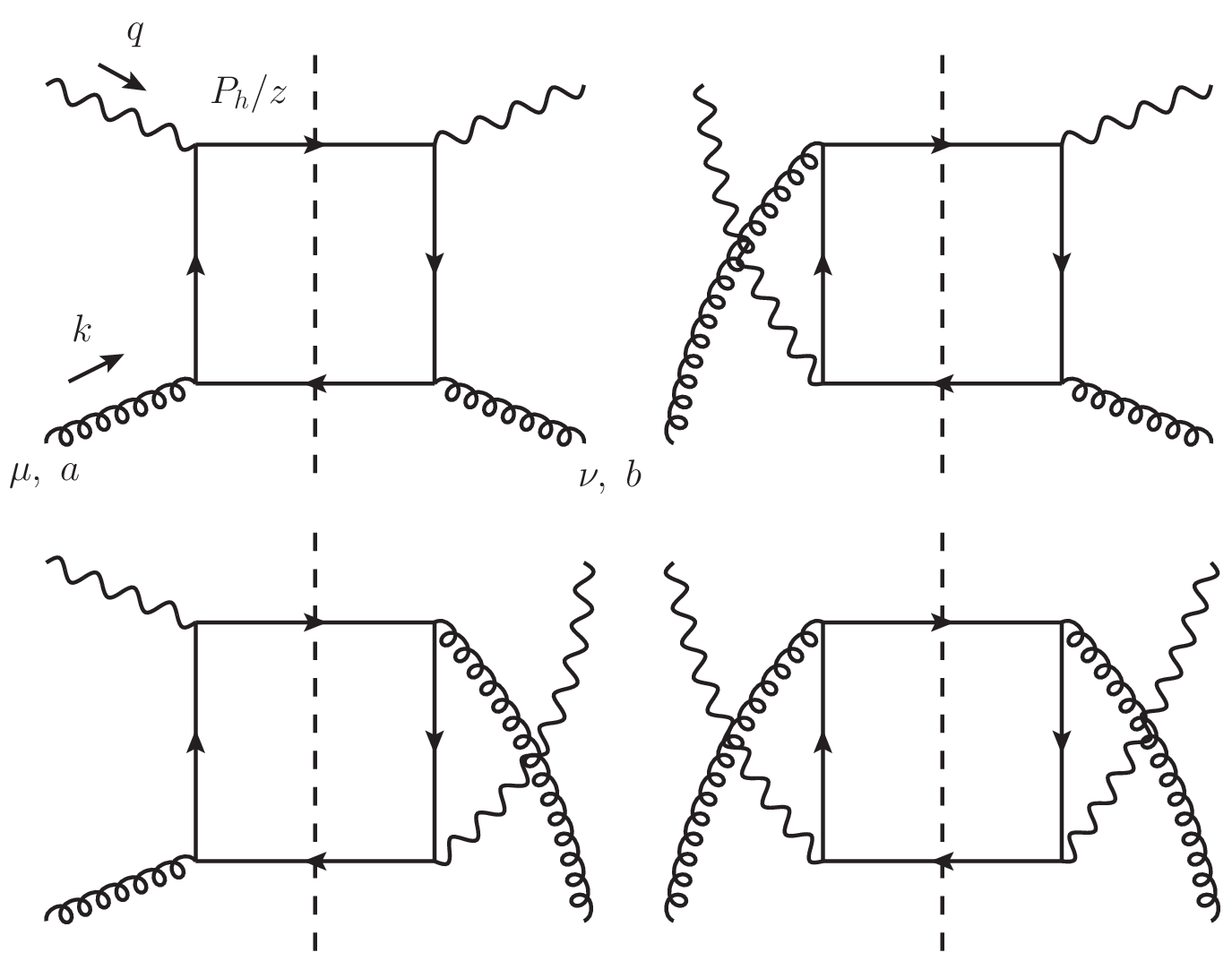}
\end{center}
 \caption{Hard scattering diagrams in $S^{ab}_{\mu\nu}(k)$.}
\end{figure}
\noindent
These diagrams satisfy the WTIs
\beq
k^{\mu}S_{\mu\nu}(k)=0,\hspace{5mm}k^{\nu}S_{\mu\nu}(k)=0,
\eeq
where we defined $S_{\mu\nu}(k)=\delta^{ab}S^{ab}_{\mu\nu}(k)/(N_c^2-1)$.
These WTIs give
\beq
p^{\mu}S^{ab}_{\mu\nu}(k)=-{1\over k\cdot n}\omega^{\mu'}_{\ \mu}
k^{\mu}S^{ab}_{\mu'\nu}(k),\hspace{5mm}
p^{\nu}S^{ab}_{\mu\nu}(k)=-{1\over k\cdot n}\omega^{\nu'}_{\ \nu}
k^{\nu}S^{ab}_{\mu\nu'}(k).
\eeq
Using these relations, we can show
\beq
W^{\rm Fig.6left}&=&\omega^{\mu'}_{\ \mu}\omega^{\nu'}_{\ \nu}
\int{d^4k\over (2\pi)^4}\int d^4\xi\,e^{ik\cdot \xi}{1\over (k\cdot n)^2}
\la P|F_a^{(0)\nu n}(0)F_a^{(0)\mu n}(\xi)|P\ra S_{\mu'\nu'}(k).
\eeq
The twist-4 contribution is given by the second derivative term of $S_{\mu'\nu'}(k)$,
\beq
&&W^{\rm Fig.6left}\Bigr|_{\rm twist-4}
\nonumber\\
&=&\omega^{\mu'}_{\ \mu}\omega^{\nu'}_{\ \nu}
\int{d^4k\over (2\pi)^4}\int d^4\xi\,e^{ik\cdot \xi}{1\over (k\cdot n)^2}
\la P|F_a^{(0)\nu n}(0)F_a^{(0)\mu n}(\xi)|P\ra 
{1\over 2}{\partial^2\over \partial k^{\alpha'}\partial k^{\beta'}}S_{\mu'\nu'}(k)\Bigr|_{k=xp}
\omega^{\alpha'}_{\ \alpha}k^{\alpha}\omega^{\beta'}_{\ \beta}k^{\beta}
\nonumber\\
&=&-{1\over 2}\omega^{\mu'}_{\ \mu}\omega^{\nu'}_{\ \nu}\omega^{\alpha'}_{\ \alpha}
\omega^{\beta'}_{\ \beta}
\int {dx\over x^2}\int {d\lambda\over 2\pi}e^{i\lambda x}
\la P|F_a^{(0)\nu n}(0)\partial^{\beta}\partial^{\alpha}F_a^{(0)\mu n}(\lambda n)|P\ra 
{\partial^2\over \partial k^{\alpha'}\partial k^{\beta'}}S_{\mu'\nu'}(k)\Bigr|_{k=xp}.
\label{Fig.6left_twist-4}
\eeq
The center diagram in Fig. \ref{gluon_diagrammatic} is given by
\beq
W^{\rm Fig.6center}={1\over 2}\int{d^4k_1\over (2\pi)^4}\int{d^4k_2\over (2\pi)^4}
\int d^4\xi\int d^4\eta\,e^{ik_1\cdot \xi}e^{i(k_2-k_1)\cdot \eta}
\la P|A_b^{\nu}(0)gA_c^{\alpha}(\eta)A_a^{\mu}(\xi)|P\ra S^{abc}_{\mu\nu\alpha}(k_1,k_2).
\eeq
The diagrams contributing to $S^{abc}_{\mu\nu\alpha}(k_1,k_2)$ are shown below.
\begin{figure}[H]
\begin{center}
  \includegraphics[height=8cm,width=12cm]{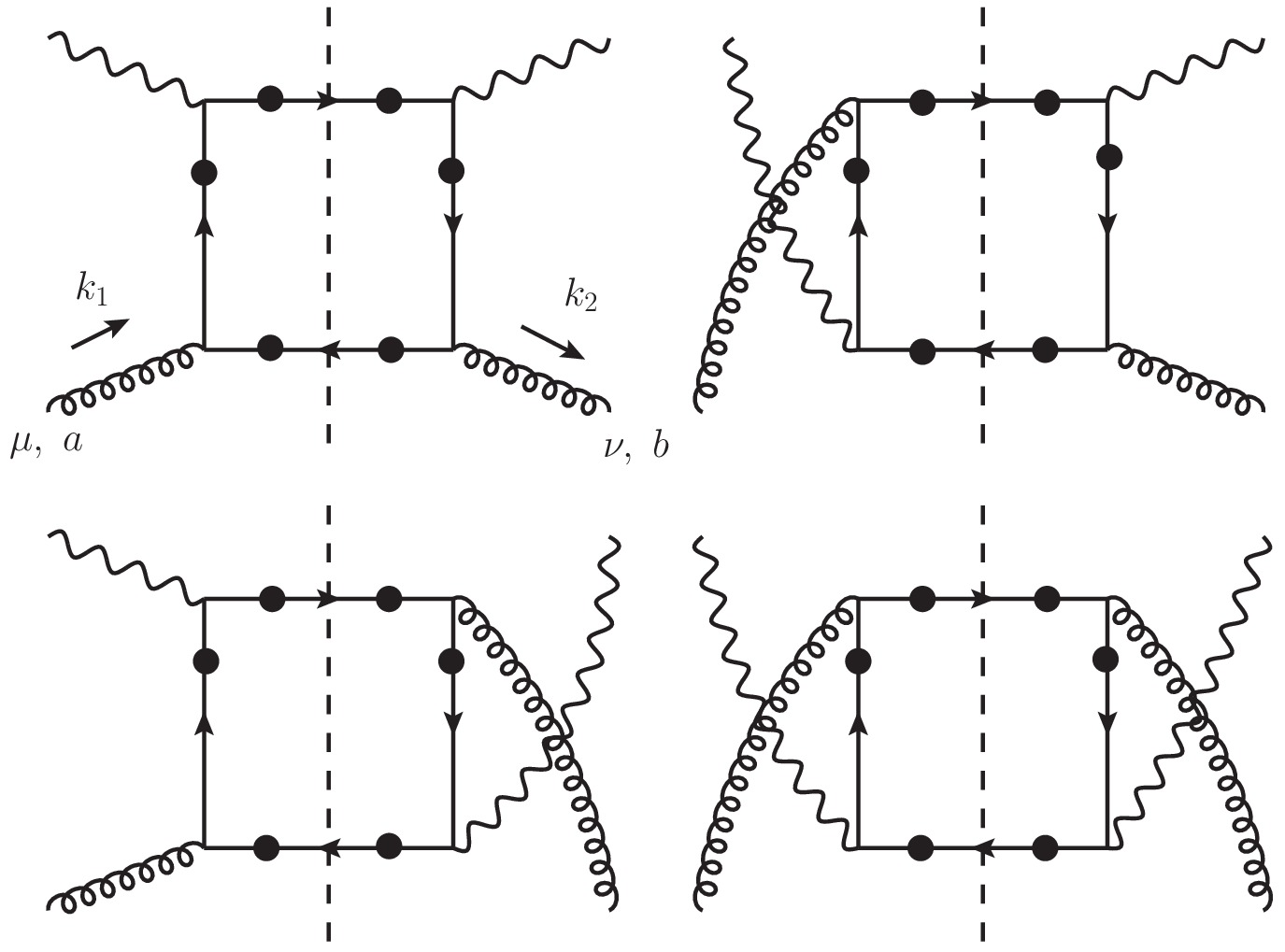}
\end{center}
 \caption{Hard scattering diagrams in $S^{abc}_{\mu\nu\alpha}(k_1,k_2)$.
 The gluon line with the momentum $k_2-k_1$ and the indices $\alpha$ and $c$ is 
 attached to one of the black dots, which means there are 24 diagrams in this figure.}
 \label{gluon_center_hard}
\end{figure}
\noindent
They satisfy the WTIs
\beq
(k_2-k_1)^{\alpha}S^{abc}_{\mu\nu\alpha}(k_1,k_2)&=&if^{abc}
\Bigl(S_{\mu\nu}(k_2)-S_{\mu\nu}(k_1)\Bigr),
\nonumber\\
k_1^{\mu}S^{abc}_{\mu\nu\alpha}(k_1,k_2)&=&-if^{abc}
S_{\alpha\nu}(k_2),
\nonumber\\
k_2^{\nu}S^{abc}_{\mu\nu\alpha}(k_1,k_2)&=&-if^{abc}
S_{\mu\alpha}(k_1),
\eeq
which give
\beq
p^{\alpha}S^{abc}_{\mu\nu\alpha}(k_1,k_2)&=&{1\over (k_2-k_1)\cdot n-i\epsilon}
\Big[-\omega^{\alpha'}_{\ \alpha}(k_2-k_1)^{\alpha}S^{abc}_{\mu\nu\alpha'}(k_1,k_2)
+if^{abc}\Bigl(S_{\mu\nu}(k_2)-S_{\mu\nu}(k_1)\Bigr)\Bigr],
\nonumber\\
p^{\mu}S^{abc}_{\mu\nu\alpha}(k_1,k_2)&=&{1\over k_1\cdot n-i\epsilon}
\Big[-\omega^{\mu'}_{\ \mu}k_1^{\mu}S^{abc}_{\mu'\nu\alpha}(k_1,k_2)
-if^{abc}S_{\alpha\nu}(k_2)\Bigr],
\nonumber\\
p^{\nu}S^{abc}_{\mu\nu\alpha}(k_1,k_2)&=&{1\over k_2\cdot n+i\epsilon}
\Big[-\omega^{\nu'}_{\ \nu}k_2^{\nu}S^{abc}_{\mu\nu'\alpha}(k_1,k_2)
-if^{abc}S_{\mu\alpha}(k_1)\Bigr].
\eeq
Using these identities, we rewrite the matrix element in the following three steps.

\

\noindent
\underline{Step 1}
\beq
A_{a}^{\mu}(\xi)S_{\mu\nu\alpha}^{abc}=
-{1\over k_1\cdot n-i\epsilon}\omega^{\mu'}_{\ \mu}
\Bigl(k_1^{\mu}A_{a}^{n}(\xi)-k_1\cdot n A_{a}^{\mu}(\xi)\Bigr)
S_{\mu'\nu\alpha}^{abc}
-{1\over k_1\cdot n- i\epsilon}A_{a}^{n}(\xi)
if^{abc}S_{\alpha\nu}(k_2),
\eeq

\noindent
\underline{Step 2}
\beq
A_{b}^{\nu}(0)S_{\mu'\nu\lambda}^{abc}&=&
-{1\over k_2\cdot n+i\epsilon}\omega^{\nu'}_{\ \nu}
\Bigl(k_2^{\nu}A_{b}^{n}(0)-k_2\cdot n A_{b}^{\nu}(0)\Bigr)
S_{\mu'\nu'\alpha}^{abc}
\nonumber\\
&&-{1\over k_2\cdot n+i\epsilon}A^n_b(0)
if^{abc}S_{\mu'\alpha}(k_1),
\eeq
\beq
&&A_{b}^{\nu}(0)S_{\alpha\nu}(k_2)=
-{1\over k_2\cdot n}\omega^{\nu'}_{\ \nu}
\Bigl(k_2^{\nu}A_{b}^{n}(0)-k_2\cdot n A_{b}^{\nu}(0)\Bigr)
S_{\alpha\nu'}(k_2),
\eeq

\begin{figure}[H]
\begin{center}
  \includegraphics[height=8cm,width=12cm]{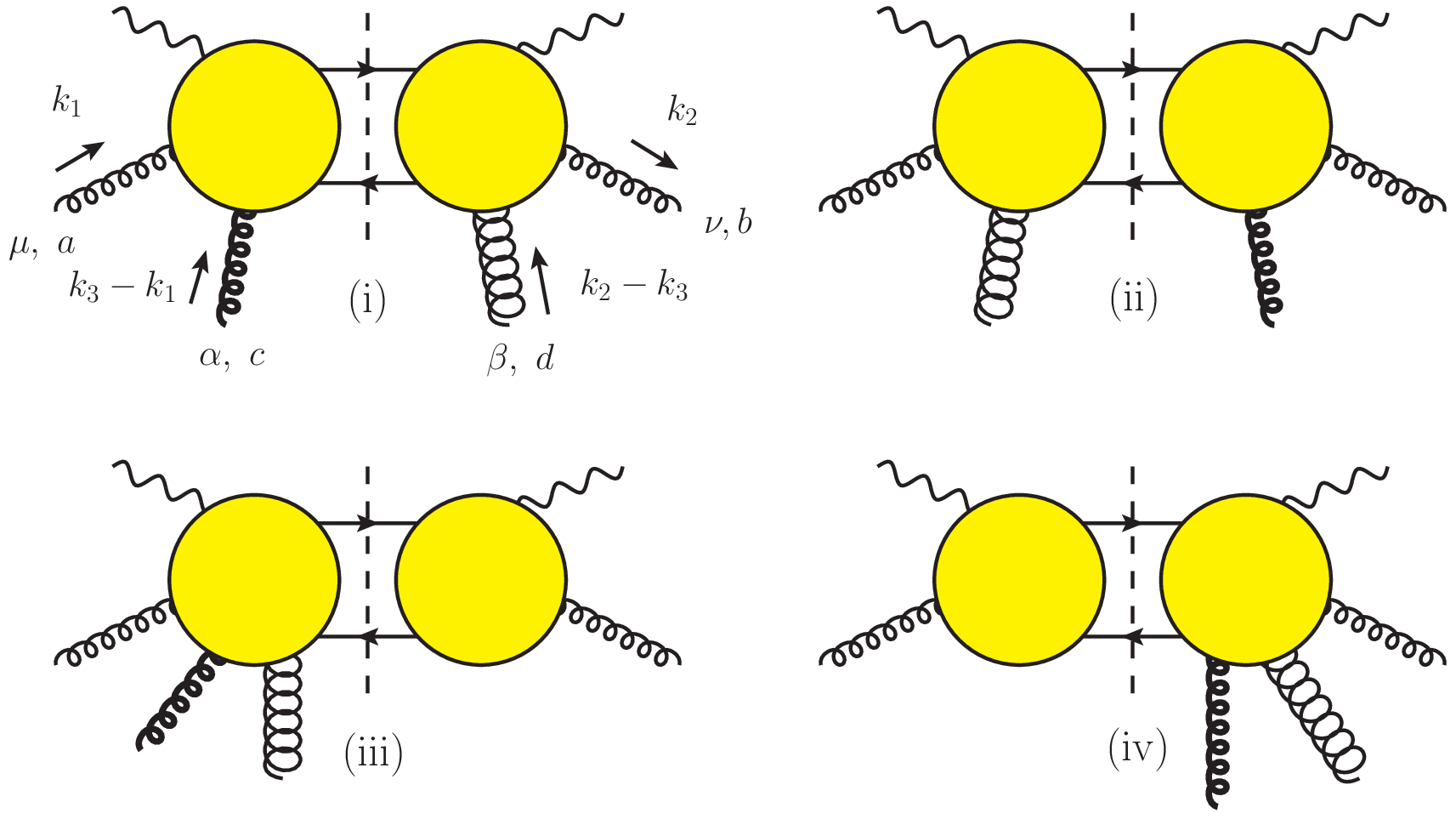}
\end{center}
 \caption{Hard scattering diagrams in $S^{({\rm i,ii,iii,iv})abcd}_{\mu\nu\alpha\beta}(k_1,k_2,k_3)$.
 The thick line represents the gluon line with momentum $k_3-k_1$ and the 
 high-amplitude line represents the gluon line with momentum $k_2-k_3$.}
 \label{gluon_right_hard}
\end{figure}

\noindent
\underline{Step 3}
\beq
A_{c}^{\alpha}(\eta)S_{\mu'\nu'\alpha}^{abc}&=&
-{1\over (k_2-k_1)\cdot n-i\epsilon}\omega^{\alpha'}_{\ \alpha}
\Bigl((k_2-k_1)^{\alpha}A_{c}^{n}(\eta)-(k_2-k_1)\cdot n A_{c}^{\alpha}(\eta)\Bigr)
S_{\mu'\nu'\alpha'}^{abc}
\nonumber\\
&&+{1\over (k_2-k_1)\cdot n- i\epsilon}A_{c}^{n}(\eta)
if^{abc}\Bigl(S_{\mu'\nu'}(k_2)-S_{\mu'\nu'}(k_1)\Bigr),
\eeq
\beq
A_{b}^{n}(0)A_{c}^{\alpha}(\eta)S_{\mu'\alpha}(k_1)
=-{1\over k_1\cdot n}\omega^{\alpha'}_{\ \alpha}
\Bigl(k_1^{\alpha}A_{b}^{n}(0)A_{c}^{n}(\eta)-k_1\cdot n A_{b}^{n}(0)A_{c}^{\alpha}(\eta)\Bigr)
S_{\mu'\alpha'}(k_1),
\eeq
\beq
A_{c}^{\alpha}(\eta)A_{a}^{n}(\xi)S_{\alpha\nu'}(k_2)
=-{1\over k_2\cdot n}\omega^{\alpha'}_{\ \alpha}
\Bigl(k_2^{\alpha}A_{c}^{n}(\eta)A_{a}^{n}(\xi)
-k_2\cdot n A_{c}^{\alpha}(\eta)A_{a}^{n}(\xi)\Bigr)
S_{\alpha'\nu'}(k_2),
\eeq
After integrating by parts, we obtain
\beq
\Bigl(k_1^{\mu}A_{a}^{n}(\xi)-k_1\cdot n A_{a}^{\mu}(\xi)\Bigr)\hspace{5mm}&\to&\hspace{5mm}
iF^{(0)\mu n}_a(\xi),
\\
\Bigl(k_2^{\nu}A_{b}^{n}(0)-k_2\cdot n A_{b}^{\nu}(0)\Bigr)\hspace{5mm}&\to&\hspace{5mm}
-iF^{(0)\nu n}_b(0),
\\
\Bigl((k_2-k_1)^{\alpha}A_{c}^{n}(\eta)-(k_2-k_1)\cdot n A_{c}^{\alpha}(\eta)\Bigr)
\hspace{5mm}&\to&\hspace{5mm}
iF^{(0)\alpha n}_c(\eta),
\\
\Bigl(k_1^{\alpha}A_{b}^{n}(0)A_{c}^{n}(\eta)-k_1\cdot n A_{b}^{n}(0)
A_{c}^{\alpha}(\eta)\Bigr)\hspace{5mm}&\to&\hspace{5mm}
-iF^{(0)\alpha n}_b(0)A^n_c(\eta)-iA^n_b(0)F^{(0)\alpha n}_c(\eta)
\\
&&+k_2\cdot n\Bigl(A^{\alpha}_b(0)A^n_c(\eta)-A^{n}_b(0)A^{\alpha}_c(\eta)\Bigr),
\\
\Bigr(k_2^{\alpha}A^n_c(\eta) A^n_a(\xi)-k_2\cdot n A^{\alpha}_c(\eta)A^{n}_a(\xi)\Bigr)
\hspace{5mm}&\to&\hspace{5mm}
iF_c^{(0)\alpha n}(\eta)A_a^n(\xi)+iA^n_c(\eta)F_a^{(0)\alpha n}(\xi)
\nonumber\\
&&\hspace{5mm}
+k_1\cdot n \Bigl(A_c^{n}(\eta)A_a^{\alpha}(\xi)-A_c^{\alpha}(\eta)A_a^{n}(\xi)\Bigr).
\eeq
We finally obtain
\beq
&&W^{\rm Fig.6center}
\nonumber\\
&=&\int{d^4k_1\over (2\pi)^4}\int{d^4k_2\over (2\pi)^4}
\int d^4\xi\int d^4\eta\,e^{ik_1\cdot \xi}e^{i(k_2-k_1)\cdot \eta}
\nonumber\\
&&\times\Bigl[
-{i\over 2}\omega^{\mu'}_{\ \mu}\omega^{\nu'}_{\ \nu}\omega^{\alpha'}_{\ \alpha}
{1\over k_1\cdot n-i\epsilon}{1\over k_2\cdot n+i\epsilon}
{1\over (k_2-k_1)\cdot n-i\epsilon}
\la P|F_b^{(0)\nu n}(0)gF_c^{(0)\alpha n}(\eta)F_a^{(0)\mu n}(\xi)|P\ra S^{abc}_{\mu'\nu'\alpha'}(k_1,k_2)
\nonumber\\
&&+\omega^{\mu'}_{\ \mu}\omega^{\nu'}_{\ \nu}
{1\over (k_2\cdot n)^2}{1\over (k_2-k_1)\cdot n-i\epsilon}
\la P|F_b^{(0)\nu n}(0)gA_c^{n}(\eta)F_a^{(0)\mu n}(\xi)|P\ra 
(T^c)_{ba}S_{\mu'\nu'}(k_2)
\nonumber\\
&&-\omega^{\mu'}_{\ \mu}\omega^{\nu'}_{\ \nu}
{1\over (k_1\cdot n)^2}{1\over (k_2-k_1)\cdot n-i\epsilon}
\la P|F_b^{(0)\nu n}(0)gA_c^{n}(\eta)F_a^{(0)\mu n}(\xi)|P\ra 
(T^c)_{ba}S_{\mu'\nu'}(k_1)
\nonumber\\
&&+\omega^{\nu'}_{\ \nu}\omega^{\alpha'}_{\ \alpha}
{1\over (k_2\cdot n)^2}
\la P|F_b^{(0)\nu n}(0)\Bigl(gf^{abc}A_c^{\alpha}(\eta)A_a^{n}(\xi)\Bigr)|P\ra S_{\alpha'\nu'}(k_2)
\nonumber\\
&&+\omega^{\mu'}_{\ \mu}\omega^{\alpha'}_{\ \alpha}
{1\over (k_1\cdot n)^2}
\la P|\Bigl(gf^{abc}A_b^{\alpha}(0)A_c^{n}(\eta)\Bigr)F_a^{(0)\mu n}(\xi)|P\ra S_{\mu'\alpha'}(k_1)
\Bigr],
\eeq
where we appropriately changed the dummy indices and the integration variables to
$k_1\to k_2-k_1$, $k_2\to k_1-k_2$ and $\xi\leftrightarrow \eta$.
The calculation is the same as that in the quark case. We pick up the first derivative terms 
of $S^{abc}_{\mu'\nu'\alpha'}(k_1,k_2)$ and the second derivative term of $S_{\mu'\nu'}(k_i)$
in the Taylor expansion,
\beq
&&W^{\rm Fig.6center}\Bigr|_{twist-4}
\nonumber\\
&=&{1\over 2}
\omega^{\mu'}_{\ \mu}\omega^{\nu'}_{\ \nu}\omega^{\alpha'}_{\ \alpha}
\omega^{\beta'}_{\ \beta}\int{dx_1}\int{dx_2}
\int {d\lambda\over 2\pi}\int {d\mu\over 2\pi}\,e^{i\lambda x_1}e^{i\mu(x_2-x_1)}
{1\over x_1-i\epsilon}{1\over x_2+i\epsilon}{1\over x_2-x_1-i\epsilon}
\nonumber\\
&&\times
\Bigl\{\la P|F_b^{(0)\nu n}(0)gF_c^{(0)\alpha n}(\mu n)
\partial^{\beta}F_a^{(0)\mu n}(\lambda n)|P\ra 
{\partial\over \partial k_1^{\beta'}}S^{abc}_{\mu'\nu'\alpha'}(k_1,k_2)\Bigr|_{k_i=x_ip}
\nonumber\\
&&+\la P|F_b^{(0)\nu n}(0)\partial^{\beta}\Bigl(gF_c^{(0)\alpha n}(\mu n)
F_a^{(0)\mu n}(\lambda n)\Bigr)|P\ra 
{\partial\over \partial k_2^{\beta'}}S^{abc}_{\mu'\nu'\alpha'}(k_1,k_2)\Bigr|_{k_i=x_ip}\Bigr\}
\nonumber\\
&&-{1\over 2}\omega^{\mu'}_{\ \mu}\omega^{\nu'}_{\ \nu}\omega^{\alpha'}_{\ \alpha}
\omega^{\beta'}_{\ \beta}\int{dx\over x^2}\int {d\lambda\over 2\pi}\,e^{i\lambda x}
\Bigl\{
\la P|F^{(0)\nu n}(0)\Bigl(ig\int^{0}_{\lambda}d\mu\,A^{n}(\mu n)\Bigr)
\partial^{\beta}\partial^{\alpha}F^{(0)\mu n}(\lambda n)|P\ra 
\nonumber\\
&&+\la P|F_b^{(0)\nu n}(0)
\partial^{\beta}\partial^{\alpha}\Bigl(gf^{bca}A_c^{\alpha}(\lambda n)A_a^{n}(\lambda n)\Bigr)|P\ra
+\la P|\Bigl(gf^{abc}A_b^{\alpha}(0)A_c^{n}(0)\Bigr)\partial^{\beta}\partial^{\alpha}
F_a^{(0)\mu n}(\lambda n)|P\ra
\nonumber\\
&&+i\int^{\infty}_{\lambda}d\mu\,\la P|F^{(0)\nu n}(0)\partial^{\beta}\Bigl(
gF^{(0)\alpha n}(\mu n)F^{(0)\mu n}(\lambda n)\Bigr)|P\ra
+\la P|F^{(0)\nu n}(0)\partial^{\beta}
\Bigl[\Bigl(-igA^{\alpha}(\lambda n)\Bigr)F^{(0)\mu n}(\lambda n)\Bigr]|P\ra
\nonumber\\
&&+i\int^{\infty}_{\lambda}d\mu\,\la P|F^{(0)\nu n}(0)gF^{(0)\beta n}(\mu n)\partial^{\alpha}
F^{(0)\mu n}(\lambda n)|P\ra
+\la P|F^{(0)\nu n}(0)\Bigl(-igA^{\beta}(\lambda n)\Bigr)\partial^{\alpha}
F^{(0)\mu n}(\lambda n)|P\ra
\nonumber\\
&&\Bigr\}{\partial^2\over \partial k_1^{\alpha'}\partial k_1^{\beta'}}S_{\mu'\nu'}(k_1)\Bigr|_{k_1=x_1p}.
\label{Fig.6center_twist-4}
\eeq
For later convenience, we expand $S^{abc}_{\mu'\nu'\alpha'}(k_1,k_2)$ in terms of 
SU($N_c$) structure constants,
\beq
S^{abc}_{\mu'\nu'\alpha'}(k_1,k_2)=if^{abc}S^{(f)}_{\mu'\nu'\alpha'}(k_1,k_2)
+d^{abc}S^{(d)}_{\mu'\nu'\alpha'}(k_1,k_2).
\eeq
Thus the relevant part in (\ref{Fig.6center_twist-4}) is rewritten as
\beq
&&\sum_{i=f,d}C_{(i)}^{abc}\Bigl[\la P|F_b^{(0)\nu n}(0)F_c^{(0)\alpha n}(\mu n)
\partial^{\beta}F_a^{(0)\mu n}(\lambda n)|P\ra 
{\partial\over \partial k_1^{\beta'}}S^{(i)}_{\mu'\nu'\alpha'}(k_1,k_2)\Bigr|_{k_i=x_ip}
\nonumber\\
&&\la P|F_b^{(0)\nu n}(0)\partial^{\beta}\Bigl(F_c^{(0)\alpha n}(\mu n)
F_a^{(0)\mu n}(\lambda n)\Bigr)|P\ra 
{\partial\over \partial k_2^{\beta'}}S^{(i)}_{\mu'\nu'\alpha'}(k_1,k_2)\Bigr|_{k_i=x_ip}\Bigr],
\label{Fig.6center_twist-4_2}
\eeq
where $C_{(f,d)}^{abc}=if^{abc},\ d^{abc}$.
We consider the right diagram in Fig. \ref{gluon_diagrammatic} which is given by
\beq
W^{\rm Fig.6right}&=&
\int{d^4k_1\over (2\pi)^4}\int{d^4k_2\over (2\pi)^4}\int{d^4k_3\over (2\pi)^4}
\int d^4\xi\int d^4\eta\int d^4\chi\,e^{ik_1\cdot \xi}e^{i(k_3-k_1)\cdot \eta}e^{i(k_2-k_3)\cdot \chi}
\nonumber\\
&&\times +\la P|A_b^{\nu}(0)gA_d^{\beta}(\chi)gA_c^{\alpha}(\eta)A_a^{\mu}(\xi)|P\ra 
\Bigl[{1\over 8}\Bigl(S^{({\rm i})abcd}_{\mu\nu\alpha\beta}(k_1,k_2,k_3)
+S^{({\rm ii})abcd}_{\mu\nu\alpha\beta}(k_1,k_2,k_3)\Bigr)
\nonumber\\
&&+{1\over 3!}\Bigl(S^{({\rm iii})abcd}_{\mu\nu\alpha\beta}(k_1,k_2,k_3)
+S^{({\rm iv})abcd}_{\mu\nu\alpha\beta}(k_1,k_2,k_3)\Bigr)
\Bigr].
\eeq
The diagrams in $S^{({\rm i,ii,iii,iv})abcd}_{\mu\nu\alpha\beta}(k_1,k_2,k_3)$ are schematically shown in 
Fig. \ref{gluon_right_hard}.
\noindent
Note that the symmetry factors are different between the pair of (i) and (ii) and 
the pair of (iii) and (iv). The former pair requires $1/(2\cdot 2\cdot 2)$
($k_1\leftrightarrow k_3-k_1$, $k_2\leftrightarrow k_2-k_3$,
$k_3-k_1\leftrightarrow k_2-k_3$), while the latter pair requires 1/(3!)
($k_{1,2}\leftrightarrow k_3-k_1\leftrightarrow k_2-k_3$).
Because of this difference, the WTIs have to be separated as follows:
\beq
(k_3-k_1)^{\alpha}\Bigl(S^{({\rm i})abcd}_{\mu\nu\alpha\beta}(k_1,k_2,k_3)
+S^{({\rm ii})abcd}_{\mu\nu\alpha\beta}(k_1,k_2,k_3)\Bigr)&=&
if^{cae}S^{ebd}_{R\mu\nu\beta}(k_3,k_2)
+if^{cbe}S^{aed}_{L\mu\nu\beta}(k_1,k_1+k_2-k_3),
\nonumber\\
(k_2-k_3)^{\beta}\Bigl(S^{({\rm i})abcd}_{\mu\nu\alpha\beta}(k_1,k_2,k_3)
+S^{({\rm ii})abcd}_{\mu\nu\alpha\beta}(k_1,k_2,k_3)\Bigr)&=&
if^{dbe}S^{aec}_{L\mu\nu\alpha}(k_1,k_3)
+if^{dae}S^{ebc}_{R\mu\nu\alpha}(k_1+k_2-k_3,k_2),
\nonumber\\
k_1^{\mu}\Bigl(S^{({\rm i})abcd}_{\mu\nu\alpha\beta}(k_1,k_2,k_3)
+S^{({\rm ii})abcd}_{\mu\nu\alpha\beta}(k_1,k_2,k_3)\Bigr)&=&
if^{ace}S^{ebd}_{R\alpha\nu\beta}(k_3,k_2)+if^{ade}S^{ebc}_{R\beta\nu\alpha}(k_1+k_2-k_3,k_2),
\nonumber\\
k_2^{\nu}\Bigl(S^{({\rm i})abcd}_{\mu\nu\alpha\beta}(k_1,k_2,k_3)
+S^{({\rm ii})abcd}_{\mu\nu\alpha\beta}(k_1,k_2,k_3)\Bigr)&=&
if^{dbe}S^{aec}_{L\mu\beta\alpha}(k_1,k_3)
+if^{cbe}S^{aed}_{L\mu\alpha\beta}(k_1,k_1+k_2-k_3).
\eeq
\beq
(k_3-k_1)^{\alpha}\Bigl(S^{({\rm iii})abcd}_{\mu\nu\alpha\beta}(k_1,k_2,k_3)
+S^{({\rm iv})abcd}_{\mu\nu\alpha\beta}(k_1,k_2,k_3)\Bigr)&=&
if^{cae}S^{ebd}_{L\mu\nu\beta}(k_3,k_2)
+if^{cbe}S^{aed}_{R\mu\nu\beta}(k_1,k_1+k_2-k_3)
\nonumber\\
&&+if^{cde}S^{abe}_{\mu\nu\beta}(k_1,k_2),
\nonumber\\
(k_2-k_3)^{\beta}\Bigl(S^{({\rm iii})abcd}_{\mu\nu\alpha\beta}(k_1,k_2,k_3)
+S^{({\rm iv})abcd}_{\mu\nu\alpha\beta}(k_1,k_2,k_3)\Bigr)&=&
if^{dbe}S^{aec}_{R\mu\nu\alpha}(k_1,k_3)
+if^{dae}S^{ebc}_{L\mu\nu\alpha}(k_1+k_2-k_3,k_2)
\nonumber\\
&&-if^{cde}S^{abe}_{\mu\nu\alpha}(k_1,k_2),
\nonumber\\
k_1^{\mu}\Bigl(S^{({\rm iii})abcd}_{\mu\nu\alpha\beta}(k_1,k_2,k_3)
+S^{({\rm iv})abcd}_{\mu\nu\alpha\beta}(k_1,k_2,k_3)\Bigr)&=&
if^{ace}S^{ebd}_{L\alpha\nu\beta}(k_3,k_2)+if^{ade}S^{ebc}_{L\beta\nu\alpha}(k_1+k_2-k_3,k_2),
\nonumber\\
k_2^{\nu}\Bigl(S^{({\rm iii})abcd}_{\mu\nu\alpha\beta}(k_1,k_2,k_3)
+S^{({\rm iv})abcd}_{\mu\nu\alpha\beta}(k_1,k_2,k_3)\Bigr)&=&
if^{dbe}S^{aec}_{R\mu\beta\alpha}(k_1,k_3)
+if^{cbe}S^{aed}_{R\mu\alpha\beta}(k_1,k_1+k_2-k_3),\hspace{3mm}
\eeq
where $S^{ebd}_{L(R)\mu\nu\beta}(k_1,k_2)$ represents the diagrams 
in which the gluon line with momentum $k_2-k_1$ is attached to the left(right) side of the cut in
Fig. \ref{gluon_center_hard}. We rewrite $W^{\rm Fig.6right}$ using these WTIs.
The calculation is tedious but straightforward.
We can show that the twist-4 contributions take the following form.
\beq
&&-{1\over 2}\omega^{\mu'}_{\ \mu}\omega^{\nu'}_{\ \nu}\omega^{\alpha'}_{\ \alpha}
\omega^{\beta'}_{\ \beta}\int {dx\over x^2}\,C^{\mu\nu\alpha\beta}_{1}(x)
{\partial^2\over \partial k^{\alpha'}\partial k^{\beta'}}S_{\mu'\nu'}(k)\Bigr|_{k=xp}
\nonumber\\
&&+{1\over 2}\omega^{\mu'}_{\ \mu}\omega^{\nu'}_{\ \nu}\omega^{\alpha'}_{\ \alpha}
\omega^{\beta'}_{\ \beta}
\int{dx_1}\int{dx_2}\,
{1\over x_1-i\epsilon}{1\over x_2+i\epsilon}{1\over x_2-x_1-i\epsilon}
\nonumber\\
&&\times\sum_{i=f,d} C^{abc}_{(i)}\Bigl[C^{\mu\nu\alpha\beta}_{2abc}(x_1,x_2)
{\partial\over \partial k_1^{\beta'}}S^{(i)}_{\mu'\nu'\alpha'}(k_1,k_2)\Bigr|_{k_i=x_ip}
+C^{\mu\nu\alpha\beta}_{3abc}(x_1,x_2)
{\partial\over \partial k_2^{\beta'}}S^{(i)}_{\mu'\nu'\alpha'}(k_1,k_2)\Bigr|_{k_i=x_ip}\Bigr]
\nonumber\\
&&-\omega^{\mu'}_{\ \mu}\omega^{\nu'}_{\ \nu}\omega^{\alpha'}_{\ \alpha}
\omega^{\beta'}_{\ \beta}\int{dx_1}\int{dx_2}\int{dx_3}\,
{1\over x_1-i\epsilon}{1\over x_2+i\epsilon}
{1\over x_3-x_1-i\epsilon}{1\over x_2-x_3-i\epsilon}
C^{\mu\nu\alpha\beta}_{4\,abcd}(x_1,x_2,x_3)
\nonumber\\
&&\times\Bigl[{1\over 8}\Bigl(S^{({\rm i})abcd}_{\mu'\nu'\alpha'\beta'}(x_1p,x_2p,x_3p)
+S^{({\rm ii})abcd}_{\mu'\nu'\alpha'\beta'}(x_1p,x_2p,x_3p)\Bigr)
\nonumber\\
&&+{1\over 3!}\Bigl(S^{({\rm iii})abcd}_{\mu'\nu'\alpha'\beta'}(x_1p,x_2p,x_3p)
+S^{({\rm iv})abcd}_{\mu'\nu'\alpha'\beta'}(x_1p,x_2p,x_3p)\Bigr)\Bigr],
\eeq
which takes a form corresponding to (\ref{twist-4_form_quark}) in the quark case.
We can extract the relevant term in $C^{\mu\nu\alpha\beta}_{4\,abcd}(x_1,x_2,x_3)$
from $W^{\rm Fig.6right}$ using WTIs, 
\beq
&&-\omega^{\mu'}_{\ \mu}\omega^{\nu'}_{\ \nu}\omega^{\alpha'}_{\ \alpha}
\omega^{\beta'}_{\ \beta}
\int{d^4k_1\over (2\pi)^4}\int{d^4k_2\over (2\pi)^4}\int{d^4k_3\over (2\pi)^4}
\int d^4\xi\int d^4\eta\int d^4\chi\,e^{ik_1\cdot \xi}e^{i(k_3-k_1)\cdot \eta}e^{i(k_2-k_3)\cdot \chi}
\nonumber\\
&&\times {1\over k_1\cdot n-i\epsilon}{1\over k_2\cdot n+i\epsilon}
{1\over (k_3-k_1)\cdot n-i\epsilon}{1\over (k_2-k_3)\cdot n-i\epsilon}
\la P|F_b^{(0)\nu n}(0)gF_d^{(0)\beta n}(\chi)gF_c^{(0)\alpha n}(\eta)F_a^{(0)\mu n}(\xi)|P\ra 
\nonumber\\
&&\times\Bigl[{1\over 8}\Bigl(S^{({\rm i})abcd}_{\mu'\nu'\alpha'\beta'}(k_1,k_2,k_3)
+S^{({\rm ii})abcd}_{\mu'\nu'\alpha'\beta'}(k_1,k_2,k_3)\Bigr)
+{1\over 3!}\Bigl(S^{({\rm iii})abcd}_{\mu'\nu'\alpha'\beta'}(k_1,k_2,k_3)
+S^{({\rm iv})abcd}_{\mu'\nu'\alpha'\beta'}(k_1,k_2,k_3)\Bigr)\Bigr].
\eeq
The twist-4 contribution is given by the leading term in the Taylor expansion of 
$S^{({\rm i,ii,iii,iv})abcd}_{\mu'\nu'\alpha'\beta'}(k_1,k_2,k_3)$
and we can show
\beq
&&C^{\mu\nu\alpha\beta}_{4\,abcd}(x_1,x_2,x_3)
\nonumber\\
&=&
\int{d\lambda\over 2\pi}\,e^{i\lambda x_1}
\int {d\mu\over 2\pi}e^{i\mu(x_3-x_1)}\int{d\omega\over 2\pi}e^{i\omega (x_2-x_3)}\,
\la P|F_b^{\nu n}(0)gF_d^{\beta n}(\omega n)
gF_c^{\alpha n}(\mu n)F_a^{\mu n}(\lambda n)|P\ra,
\eeq
which gives (\ref{gluon_FF}) when it is contracted with the color factor $if^{cea}if^{dbe}$
in $S^{({\rm i,ii,iii,iv})abcd}_{\mu'\nu'\alpha'\beta'}(k_1,k_2,k_3)$.
The relevant terms in $C^{\mu\nu\alpha\beta}_{2,3\,abc}(x_1,x_2)$
from $W^{\rm Fig.6right}$ are given by
\beq
&&-{i\over 2}\omega^{\mu'}_{\ \mu}\omega^{\nu'}_{\ \nu}\omega^{\alpha'}_{\ \alpha}
\int{d^4k_1\over (2\pi)^4}\int{d^4k_2\over (2\pi)^4}\int{d^4k_3\over (2\pi)^4}
\int d^4\xi\int d^4\eta\int d^4\chi\,e^{ik_1\cdot \xi}e^{i(k_3-k_1)\cdot \eta}e^{i(k_2-k_3)\cdot \chi}
\nonumber\\
&&\times \Bigl\{{if^{dbe}\over k_1\cdot n-i\epsilon}{1\over k_3\cdot n+i\epsilon}
{1\over (k_3-k_1)\cdot n-i\epsilon}{1\over (k_2-k_3)\cdot n-i\epsilon}
\la P|F_b^{(0)\nu n}(0)gA_d^{n}(\chi)gF_c^{(0)\alpha n}(\eta)F_a^{(0)\mu n}(\xi)|P\ra 
S^{aec}_{\mu'\nu'\alpha'}(k_1,k_3)
\nonumber\\
&&-{if^{ace}\over k_3\cdot n-i\epsilon}{1\over k_2\cdot n+i\epsilon}
{1\over (k_3-k_1)\cdot n-i\epsilon}{1\over (k_2-k_3)\cdot n-i\epsilon}
\la P|F_b^{(0)\nu n}(0)gF_d^{(0)\alpha n}(\chi)gA_c^{n}(\eta)F_a^{(0)\mu n}(\xi)|P\ra 
S^{ebd}_{\mu'\nu'\alpha'}(k_3,k_2)
\nonumber\\
&&-{if^{cde}\over k_1\cdot n-i\epsilon}{1\over k_2\cdot n+i\epsilon}
{1\over (k_2-k_3)\cdot n-i\epsilon}{1\over (k_2-k_1)\cdot n-i\epsilon}
\la P|F_b^{(0)\nu n}(0)gA_d^{n}(\chi)gF_c^{(0)\alpha n}(\eta)F_a^{(0)\mu n}(\xi)|P\ra 
S^{abe}_{\mu'\nu'\alpha'}(k_1,k_2)\Bigr\}
\nonumber\\
&&-{i\over 2}\omega^{\mu'}_{\ \mu}\omega^{\nu'}_{\ \nu}\omega^{\alpha'}_{\ \alpha}
\int{d^4k_1\over (2\pi)^4}\int{d^4k_2\over (2\pi)^4}
\int d^4\xi\int d^4\eta\,e^{ik_1\cdot \xi}e^{i(k_2-k_1)\cdot \eta}
{1\over k_1\cdot n-i\epsilon}
{1\over k_2\cdot n+i\epsilon}{1\over (k_2-k_1)\cdot n-i\epsilon}
\nonumber\\
&&\times \Bigl\{
\la P|\Bigl(gf^{ebd}A_b^{\nu}(0)A_d^{n}(0)\Bigr)
gF_c^{(0)\alpha n}(\eta)F_a^{(0)\mu n}(\xi)|P\ra 
S^{aec}_{\mu'\nu'\alpha'}(k_1,k_2)
\nonumber\\
&&+\la P|F_b^{(0)\nu n}(0)gF_d^{(0)\alpha n}(\eta)
\Bigl(gf^{ace}A_a^{\mu}(\xi)A_c^{n}(\xi)\Bigr)|P\ra 
S^{ebd}_{\mu'\nu'\alpha'}(k_1,k_2)
\nonumber\\
&&+\la P|F_b^{(0)\nu n}(0)\Bigl(gf^{dce}A_d^{\alpha}(\eta)
A_c^{n}(\eta)\Bigr)gF_a^{(0)\mu n}(\xi)|P\ra 
S^{abe}_{\mu'\nu'\alpha'}(k_1,k_2)
\Bigr\},
\eeq
where one may use symmetries of gluon external lines,
\beq
S^{ebd}_{L\mu'\nu'\alpha'}(k_3,k_2)=S^{dbe}_{L\alpha'\nu'\mu'}(k_2-k_3,k_2),\hspace{5mm}
S^{aec}_{R\mu'\nu'\alpha'}(k_1,k_3)=S^{ace}_{R\mu'\alpha'\nu'}(k_1,k_1-k_3).
\eeq
Combining the first derivative terms of $S^{(f,d)}_{\mu'\nu'\alpha'}(k_i,k_j)$ in the Taylor expansion
and (\ref{Fig.6center_twist-4_2}), we can show
\beq
C^{\mu\nu\alpha\beta}_{2\,abc}(x_1,x_2)=
M^{G\,\mu\nu\alpha\beta}_{1\partial\,abc}(x_1,x_2)\Bigr|_{{\rm up\ to\ }g^2},
\hspace{5mm}
C^{\mu\nu\alpha\beta}_{3\,abc}(x_1,x_2)=
M^{G\,\mu\nu\alpha\beta}_{2\partial\,abc}(x_1,x_2)\Bigr|_{{\rm up\ to\ }g^2}.
\eeq
These matrix elements give (\ref{gluon_kinematical_1}) and (\ref{gluon_kinematical_2})
when they are contracted with $if^{abc}$.
The relevant terms in $C^{\mu\nu\alpha\beta}_{1}(x)$
from $W^{\rm Fig.6right}$ are given by
\beq
&&\omega^{\mu'}_{\ \mu}\omega^{\nu'}_{\ \nu}
\int{d^4k_1\over (2\pi)^4}\int{d^4k_2\over (2\pi)^4}\int{d^4k_3\over (2\pi)^4}
\int d^4\xi\int d^4\eta\int d^4\chi\,e^{ik_1\cdot \xi}e^{i(k_3-k_1)\cdot \eta}e^{i(k_2-k_3)\cdot \chi}
\nonumber\\
&&\times \la P|F_b^{(0)\nu n}(0)gA_d^{n}(\chi)gA_c^{n}(\eta)F_a^{(0)\mu n}(\xi)|P\ra 
\Bigl\{\Bigl({if^{ace}if^{dbe}\over (k_1\cdot n)^2}
{1\over (k_3-k_1)\cdot n-i\epsilon}{1\over (k_2-k_1)\cdot n-i\epsilon}
S_{\mu'\nu'}(k_1)
\nonumber\\
&&+{if^{ace}if^{dbe}\over (k_2\cdot n)^2}
{1\over (k_2-k_3)\cdot n-i\epsilon}{1\over (k_2-k_1)\cdot n-i\epsilon}S_{\mu'\nu'}(k_2)
-{if^{ace}if^{dbe}\over (k_3\cdot n)^2}
{1\over (k_3-k_1)\cdot n-i\epsilon}{1\over (k_2-k_3)\cdot n-i\epsilon}S_{\mu'\nu'}(k_3)
\Bigr\}
\nonumber\\
&&+\omega^{\mu'}_{\ \mu}\omega^{\nu'}_{\ \nu}
\int{d^4k_1\over (2\pi)^4}\int{d^4k_3\over (2\pi)^4}
\int d^4\xi\int d^4\eta,e^{ik_1\cdot \xi}e^{i(k_3-k_1)\cdot \eta}
\la P|\Bigl(gf^{ebd}A_b^{\nu}(0)A_d^{n}(0)\Bigr)gA_c^{n}(\eta)F_a^{(0)\mu n}(\xi)|P\ra 
\nonumber\\
&&\times\Bigl\{{if^{ace}\over (k_1\cdot n)^2}
{1\over (k_3-k_1)\cdot n-i\epsilon}S_{\mu'\nu'}(k_1)
-{if^{ace}\over (k_3\cdot n)^2}{1\over (k_3-k_1)\cdot n-i\epsilon}S_{\mu'\nu'}(k_3)
\Bigr\}
\nonumber\\
&&+\omega^{\mu'}_{\ \mu}\omega^{\nu'}_{\ \nu}
\int{d^4k_2\over (2\pi)^4}\int{d^4k_3\over (2\pi)^4}
\int d^4\xi\int d^4\chi\,e^{ik_3\cdot \xi}e^{i(k_2-k_3)\cdot \chi}
\la P|F_b^{(0)\nu n}(0)gA_d^{n}(\chi)\Bigl(gf^{eac}A_a^{\mu}(\xi)A_c^{n}(\xi)\Bigr)|P\ra 
\nonumber\\
&&\times \Bigl\{
-{if^{dbe}\over (k_2\cdot n)^2}{1\over (k_2-k_3)\cdot n-i\epsilon}S_{\mu'\nu'}(k_2)
+{if^{dbe}\over (k_3\cdot n)^2}{1\over (k_2-k_3)\cdot n-i\epsilon}S_{\mu'\nu'}(k_3)
\Bigr\}
\nonumber\\
&&+\int{d^4k_3\over (2\pi)^4}\int d^4\eta\,e^{ik_3\cdot \eta}
\Bigl\{{1\over (k_3\cdot n)^2}
\omega^{\mu'}_{\ \mu}\omega^{\nu'}_{\ \nu}
\la P|\Bigl(gf^{bde}A_b^{\nu}(0)A_d^{n}(0)\Bigr)\Bigl(gf^{ace}A_a^{\mu}(\xi)A_c^{n}(\xi)\Bigr)|P\ra 
S_{\mu'\nu'}(k_3)\Bigr\}.
\eeq
Combining the second derivative terms of $S_{\mu'\nu'}(k_i)$ in the Taylor expansion, 
(\ref{Fig.6left_twist-4}) and (\ref{Fig.6center_twist-4}), we can show
\beq
C^{\mu\nu\alpha\beta}_{1}(x)=M^{G\,\mu\nu\alpha\beta}_{\partial\partial}(x)\Bigr|_{{\rm up\ to\ }g^2}.
\eeq
We finally obtain the result of gluon twist-4 contributions to the unpolarized cross section,
\beq
&&W(P,q,{P_h\over z})=-{1\over 2}\omega^{\mu'}_{\ \mu}\omega^{\nu'}_{\ \nu}\omega^{\alpha'}_{\ \alpha}
\omega^{\beta'}_{\ \beta}\int {dx\over x^2}M^{G\,\mu\nu\alpha\beta}_{\partial\partial}(x)
{\partial^2\over \partial k^{\alpha'}\partial k^{\beta'}}S_{\mu'\nu'}(k)\Bigr|_{k=xp}
\nonumber\\
&&+{1\over 2}\omega^{\mu'}_{\ \mu}\omega^{\nu'}_{\ \nu}\omega^{\alpha'}_{\ \alpha}
\omega^{\beta'}_{\ \beta}
\int{dx_1}\int{dx_2}\,
{1\over x_1-i\epsilon}{1\over x_2+i\epsilon}{1\over x_2-x_1-i\epsilon}
\nonumber\\
&&\times\sum_{i=f,d} C^{abc}_{(i)}\Bigl[M^{G\,\mu\nu\alpha\beta}_{1\partial\,abc}(x_1,x_2)
{\partial\over \partial k_1^{\beta'}}S^{(i)}_{\mu'\nu'\alpha'}(k_1,k_2)\Bigr|_{k_i=x_ip}
+M^{G\,\mu\nu\alpha\beta}_{2\partial\,abc}(x_1,x_2)
{\partial\over \partial k_2^{\beta'}}S^{(i)}_{\mu'\nu'\alpha'}(k_1,k_2)\Bigr|_{k_i=x_ip}\Bigr]
\nonumber\\
&&-\omega^{\mu'}_{\ \mu}\omega^{\nu'}_{\ \nu}\omega^{\alpha'}_{\ \alpha}
\omega^{\beta'}_{\ \beta}\int{dx_1}\int{dx_2}\int{dx_3}\,
{1\over x_1-i\epsilon}{1\over x_2+i\epsilon}
{1\over x_3-x_1-i\epsilon}{1\over x_2-x_3-i\epsilon}
\nonumber\\
&&\times \sum M^{G\,\mu\nu\alpha\beta}_{FF}(x_1,x_2,x_3)
\Bigl[{1\over 8}\Bigl(S^{({\rm i})}_{\mu'\nu'\alpha'\beta'}(x_1p,x_2p,x_3p)
+S^{({\rm ii})}_{\mu'\nu'\alpha'\beta'}(x_1p,x_2p,x_3p)\Bigr)
\nonumber\\
&&+{1\over 3!}\Bigl(S^{({\rm iii})}_{\mu'\nu'\alpha'\beta'}(x_1p,x_2p,x_3p)
+S^{({\rm iv})}_{\mu'\nu'\alpha'\beta'}(x_1p,x_2p,x_3p)\Bigr)\Bigr],
\eeq
where $\sum M^{G\,\mu\nu\alpha\beta}_{FF}(x_1,x_2,x_3)...$ takes summation over 
all independent gauge invariant matrix elements composed of four field strength tensors and
$\delta^{ab}$,  $if^{abc}$ and $d^{abc}$. Many of the matrix elements in this formula
appear in the gluon condensate term. The 4-gluon contributions were also discussed 
in Ref. \cite{Bartels:1999km}. 
However, the earlier formulation overlooked the $-i\epsilon$ prescription, and the TMD-related matrix 
elements were not constructed. Our result updates the earlier formulation by incorporating modern 
insights gained from recent developments in the twist-3 framework.
Exploring the possibility of extracting information on the gluon condensate from unpolarized 
cross sections in inclusive processes is one avenue worth pursuing.


\section{Summary}

In this paper, we explored the possibility of evaluating the gluon condensate term, which is one of the contributions to the origin of the nucleon mass, using high-energy processes. We first expressed the gluon condensate term as an expansion in terms of experimentally accessible
twist-4 matrix elements constructed from four  operators. Our result provides a basis for evaluating the gluon condensate term in terms of twist-4 functions. Furthermore, the fact that some of these matrix elements can be expressed in terms of TMD matrix elements suggests the possibility of indirectly evaluating the gluon condensate term using lower-twist TMD functions.

We also derived complete results for the twist-4 contributions to the cross section in inclusive processes including all pure quark, quark-gluon, and pure gluon contributions. Our results show that all the twist-4 matrix elements appearing in the gluon condensate term also appear in the twist-4 contribution to the cross section. Since our derivation is based only on the Ward-Takahashi identities which are generally satisfied in QCD hard processes, our method can be applied to general processes in $ep$ collisions.
Although our discussion in this paper focused on the nucleon mass, our results can also provide a basis for studying twist-4 effects in other research subjects. We hope that future studies will propose processes suitable for determining twist-4 functions at the EIC. For actual cross-section calculations, it is necessary to derive parametrizations of the individual matrix elements and relations among the functions 
associated with the equations of motion and translational invariance in order to clarify a complete set of functions at the twist-4 level. 
Thus, there still remain many issues to be studied in this direction.
We hope that our results will stimulate further discussion of an approach to the origin of the nucleon mass based on twist-4 effects toward the EIC.


\appendix

\end{document}